\documentclass[twocolumn,showpacs,pra,superscriptaddress]{revtex4}
\usepackage{dcolumn}
\usepackage{graphicx}
\usepackage{bbm}
\usepackage{amssymb,amsmath}
\usepackage{color}
\usepackage{ulem}

\newcommand{\be}{\begin{equation}}
\newcommand{\ee}{\end{equation}}
\newcommand{\beq}{\begin{eqnarray}}
\newcommand{\eeq}{\end{eqnarray}}

\newcommand{\hide}[1]{}

\newcommand{\ket}[1]{\left| #1 \right\rangle}
\newcommand{\bra}[1]{\left\langle #1 \right|}

\begin{document}

\title{Rydberg atom mediated non-destructive readout of collective rotational states in polar molecule arrays}

\author{Elena Kuznetsova}
\address{Institute of Applied Physics RAS, 46 Ulyanov Street, Nizhny Novgorod, 603950, Russia}
\address{Rzhanov Institute of Semiconductor Physics, Novosibirsk, 630090, Russia}
\address{Russian Quantum Center, 100 Novaya Street, Skolkovo, Moscow region, 143025, Russia}
\author{Seth T. Rittenhouse}
\address{Department of Physics, The United States Naval Academy, Annapolis, MD 21402, USA}
\author{H. R. Sadeghpour}
\address{ITAMP, Harvard-Smithsonian Center for Astrophysics, 60 Garden Street, Cambridge, MA 02138, USA}
\author{Susanne F. Yelin}
\address{ITAMP, Harvard-Smithsonian Center for Astrophysics, 60 Garden Street, Cambridge, MA 02138, USA}
\address{Department of Physics, University of Connecticut, 2152 Hillside Road, Storrs, CT 06269}
\address{Department of Physics, Harvard University, 17 Oxford Street, Cambridge, MA 02138, USA}

\date{\today}

\begin{abstract}

We analyze in detail the possibility to use charge-dipole interaction between a single polar molecule or a 1D molecular array and a 
single Rydberg atom to read out rotational populations. The change in the Rydberg electron energy is conditioned on the rotational 
state of the polar molecules, allowing for realization of a CNOT quantum gate between the molecules and the atom. 
Subsequent readout of the atomic fluorescence results in a non-destructive measurement of the rotational state.
We study the interaction between a 1D array of polar molecules and an array or a cloud of atoms in a Rydberg superatom (blockaded) 
state and calculate the resolved energy shifts of Rb(60s) with 
KRb and RbYb molecules, with N=1, 3, 5 molecules. We show that the collective molecular rotational states can be read out using the conditioned 
Rydberg 
energy shifts.

\end{abstract}

\maketitle

\section{Introduction}

Ultracold polar molecules placed in a periodic array represent an attractive setup for quantum computation \cite{QC-pol-mol} and simulation of 
strongly correlated manybody systems due to the ability to interact via anisotropic and long-range electric dipole-dipole interaction. Polar
molecules in optical lattices can be used to simulate quantum magnetism \cite{Pol-mol-quant-magn}, exotic topological states \cite{Pol-mol-topology}
and complex many-body entanglement \cite{Pol-mol-entanglement}. The first experimental observation of spin exchange processes
between KRb molecules in a 3D optical lattice induced by the dipole-dipole interaction has been reported recently \cite{Jin-pol-mol-spin-exchange}. 
A similar effect was also observed in a Cr gas, where the spin exchange between magnetic dipoles was induced
by magnetic dipole-dipole interaction \cite{Magnetic-spin-exchange}. Typically a spin-1/2 particle or a qubit is encoded in two rotational 
molecular states and an initial many-body state becomes strongly entangled due 
to the interaction. The state of such entanglement will need to be read out to extract useful information about the system.
One main challenge for the current ultracold molecule setups is that there 
exists no reliable scheme for readout which does not lead to loss of molecules.

In this work, we propose an elegant approach to non-destructively read out the rotational excitations in a mesoscopic ensemble 
of molecular array. 
We consider a linear or a ring 1D array of molecules interacting with a 1D array or cloud of neutral atoms in a symmetric state with a single Rydberg
excitation (superatom). We show that in this setup it is possible to measure total populations of collective rotational states.

 Most current 
methods for molecular state readout, such as inverse STIRAP combined with Feshbach dissociation for alkali dimers \cite{Jin-pol-mol-spin-exchange} 
and REMPI \cite{REMPI}, are destructive. In previous work \cite{PCCP} we proposed a technique to reading out populations of rotational 
molecular states relying on the interaction of a polar molecule with a nearby Rydberg atom, via the Rydberg electron- molecular 
dipole interaction \cite{Seth-Hossein-PRL,Seth-JPB}. This interaction 
shifts the states of the combined molecule-Rydberg atom system depending on the rotational state, which allows to measure its population
by conditionally exciting the atom to a Rydberg state and measuring atomic fluorescence intensity. The readout, therefore, does not require the 
molecule to be destroyed or lost.

Precision spectroscopy of Rydberg atoms as interacting auxiliary systems for readout of molecular states, 
which can not be easily measured due to a lack of cycling transition, is akin to 
the technique of quantum logic spectroscopy
\cite{Quantum-logic-spectroscopy}. In an earlier proposal \cite{Garcia-Ripoll-pol-mol-ion} an ion was considered for 
manipulation of internal states of a polar molecule via excitation of common mechanical oscillation modes. 
In our setup a Rydberg atom and a polar molecule are both neutral, which makes their co-trapping
and manipulation with external electric fields easier. The interaction also allows to make quantum non-demolition measurements by 
entangling the atom and the molecule \cite{QND}. 

There is strong and growing interest in manipulating states of few- to many-body systems using their interaction with a single ancilla system. 
Examples include control of environment nuclear spins by an electronic spin in an NV center in diamond \cite{N-V-nuclear-spin-control} and in a 
quantum dot \cite{QDot-nuclear-spin-control}, including electron spin mediated nuclear spin polarization \cite{Nuclear-spin-polarization}, superradiance \cite{Nuclear-spin-superradiance} and squeezing \cite{N-V-nuclear-spin-squeezing}. 
The sensitivity of precision measurement of a magnetic field by an NV center can be increased up to the Heisenberg limit using its electron 
spin interaction with surrounding nuclear spins \cite{Paola-N-V-envir-assist-measurement}. Recently these ideas have been extended to atomic 
systems as well. In \cite{Pfau-Rydb-BEC-spectroscopy} a local density of a quantum gas of ground state atoms was measured
by exciting one of the atoms to a Rydberg state and measuring its density-dependent energy shift. A single Rydberg atom in a cloud of ultracold ground state
atoms can form bound states with a mesoscopic number of atoms (up to 5 have been detected in \cite{Rydb-mol-spectra-Pfau}), and the corresponding
bound states were observed as distinct narrow peaks in Rydberg atom absorption spectra. In a Rydberg quantum simulator a single Rydberg atom can be used
for manipulation and measurement of atomic qubits on plaquettes in an optical lattice \cite{Rydb-quant-simulator}.

The paper is organized as follows. In Section II we derive matrix
elements of the Hamiltonian for the combined single molecule-single Rydberg atom system. In Section III we numerically calculate energy 
shifts of the states of KRb and RbYb molecules interacting with Rb(60s) atom. In Section IV we analyze the interaction between a linear or a 
ring 1D array of molecules and a Rydberg superatom,
placed either in a parallel commensurate 1D array or a dipole trap. In
Section V we discuss readout of rotational states of a single molecule or an array of molecules using its interaction with a Rydberg atom. 
Finally, we conclude in Section VI.

\section{Single atom - single molecule interaction}

\begin{figure}
\center{
\includegraphics[width=8.cm]{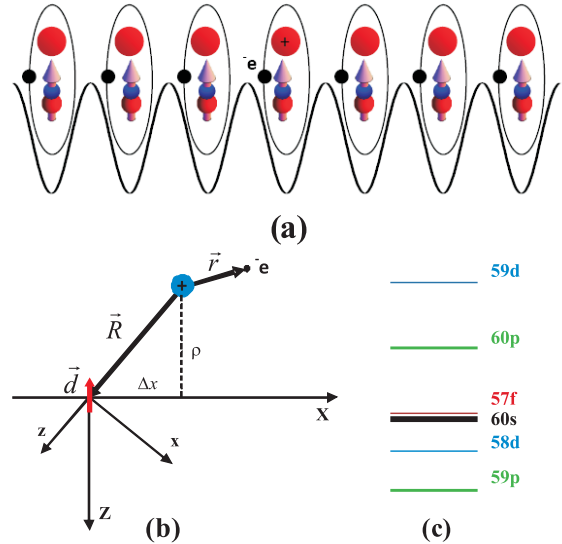}
\caption{\label{fig:setup} (a) Setup geometry: a 1D or 2D array of polar molecules, interacting via charge-dipole interaction with neaby Rydberg atoms, placed in a parallel array; (b) 
a polar molecule with a dipole moment $\vec{d}$ interacts with a Rydberg atom. The distance between the 
X axis and the Rydberg core is $\rho$, the distance between the molecule and the vertical line connecting the core and the X-axis is $\Delta x$. 
In order to calculate the energy shifts of the states of the atom-molecule system it is convenient to make a transformation to a new coordinate system 
(x,y,z) such that the z axis lies along $\vec{R}$; (c) Level scheme of Rydberg states of Rb near 60s state, taken into account in the calculations. The energy 
splittings are: $E_{60p}-E_{60s}=17.06$ GHz, $E_{60s}-E_{59p}=18.75$ GHz, $E_{59d}-E_{60s}=27.46$ GHz, $E_{60s}-E_{58d}=7.79$ GHz. }
}
\end{figure}

We envisage a setup shown in Fig.\ref{fig:setup}a where a 1D or 2D array of polar molecules is used to simulate a strongly correlated many-body quantum system. Each
polar molecule represents a qubit or a spin-1/2, encoded in rotational states $\ket{\downarrow}=\ket{J=0,m_{J}=0}$ and
$\ket{\uparrow}=\ket{J=1,m_{J}=0}$  or $\ket{\uparrow}=\ket{J=1,m_{J}=\pm 1}$ \cite{Jin-pol-mol-spin-exchange}. Parallel to the molecular array 
there is an array with neutral atoms, which can be individually excited to Rydberg states to read out molecular states.
Although a setup with two close lattices, one filled with molecules and another with atoms has yet to be realized, two parallel optical lattices filled with neutral atoms have
been demonstrated recently \cite{Greiner-parallel-lattices}.

The configuration for a single polar molecule interacting with a single Rydberg atom \cite{Seth-Hossein-PRL} is depicted in Fig.\ref{fig:setup}b:
the molecule is a part of e.g. an array aligned along the X axis with its own Rydberg atom at a distance $\rho$ from
the X axis, placed at $\Delta x= 0$.

The Hamiltonian governing the system is given by 
\begin{eqnarray}
\label{eq:hamiltonian}
H=H_{a}+H_{m}+V_{\rm e^--M},
\end{eqnarray}
where
\begin{equation}
H_{a}=\sum_{n,l,m}E_{nl}\ket{nlm}\bra{nlm} \nonumber
\end{equation}
is the unperturbed Rydberg atom Hamiltonian at principal quantum numbers $n$, orbital angular momentum $l$, and a $z$ projection of 
$l$, $m$. Electron spin mixing due to Rydberg electron spin-orbit and ground electron hyperfine interactions are not included \cite{Rydb-fine-structure}.
We calculate interaction induced energy shifts of $ns$ Rydberg states due to $V_{\rm e^--M}$ by including the $ns$ and the nearest $p$ and 
$d$, and $f$ states, whose quantum defects are nonzero: $\ket{np}$ and $\ket{(n-1)p}$, $\ket{(n-1)d}$ and $\ket{(n-2)d}$, and $\ket{(n-3)f}$. 
The corresponding unperturbed energies of Rydberg states are $E_{nl}=-1/2(n-\mu_{l})^{2}$ in atomic units, $\mu_{l}$ is the quantum defect 
(for Rb $\mu_{s}=3.13$, $\mu_{p}=2.65$, $\mu_{d}=1.34$, $\mu_{f}=0.016$).
In particular, we use $\ket{ns}=\ket{60s}$, $\ket{np}=\ket{60p}$, $\ket{(n-1)p}=\ket{59p}$, $\ket{(n-1)d}=\ket{59d}$, $\ket{(n-2)d}=\ket{58d}$ and 
$\ket{(n-3)f}=\ket{(ns-3)f}=\ket{57f}$ states. The corresponding atomic level scheme with energy splittings is shown in 
Fig.\ref{fig:setup}c. The $H_{m}=B{\bf J}^{2}$ Hamitonian
describes a rigid rotor molecule with states $\ket{J,m_{J}}$ with $H_{m}|J,m_{J}\rangle =BJ(J+1)|J,m_{J}\rangle$ and rotational 
constant $B$, and 
$V_{\rm e^--M}=\frac{e \vec{d}\cdot\vec{R}}{R^{3}}-\frac{e\vec{d}\cdot\left(\vec{R}-\vec{r}\right)}{\left|\vec{R}-\vec{r}\right|^{3}}$ 
is the charge-dipole interaction between the Rydberg atom ionic core and electron and the molecule, 
where $\vec{d}$ is the molecular dipole moment, 
$\vec{R}$ is the distance between the Rydberg core and the molecule, and $\vec{r}$ is the distance between the core and the Rydberg electron. 
The molecular permanent dipole moment is chosen to be below the Fermi-Teller critical value of $d_{cr}=1.63$ D \cite{Seth-Hossein-PRL}.

Next we calculate the shifts of the states $\ket{nlm}\ket{\downarrow}=\ket{nlm}\ket{J=0,m_{J}=0}$ and 
$\ket{nlm}\ket{\uparrow}=\ket{nlm}\ket{J=1,m_{J}=0,\pm1}$ of the combined atom-molecule system. For that we additionally take into account 
 unperturbed states 
 $\ket{nlm}\ket{J=2,m_{J}=0,\pm 1,\pm 2}$. 
The matrix elements of $H_{a}$, $H_{m}$ and $V_{\rm e^--M}$ are given by: 
\begin{eqnarray}
\label{eq:matrix-elements}
\lefteqn{\bra{J,m_{J}}\bra{n\;l\;m}H_{a}\ket{n'\;l'\;m'}\ket{J',m_{J}'} \;=}  \\
&=& -\frac{1}{2(n-\mu_{l})^{2}}\delta_{n,n'}\delta_{l,l'}\delta_{m,m'}\delta_{J,J'}\delta_{m_{J},m_{J}'}, \nonumber \\[.2cm]
\lefteqn{\bra{J,m_{J}}\bra{n\;l\;m}H_{m}\ket{n'\;l'\;m'}\ket{J',m_{J}'} \;=} \nonumber \\
&=& BJ(J+1)\delta_{n,n'}\delta_{l,l'}\delta_{m,m'}\delta_{J,J'}\delta_{m_{J},m_{J}'}, \nonumber \\[.2cm]
\lefteqn{\bra{J,m_{J}}\bra{n\;l\;m}V_{\rm e^--M}\ket{n'\;l'\;m'}\ket{J',m_{J}'}  \;=}  \nonumber \\
&=&\bra{J,m_{J}}e \vec{d}\cdot\ket{J',m_{J}'}\frac{\vec{R}}{R^{3}}\delta_{n,n'}\delta_{l,l'}\delta_{m,m'}  \nonumber \\
&&-\bra{J,m_{J}} e\vec{d} \cdot\ket{J',m_{J}'}\bra{n\;l\;m}\frac{\vec{R}-\vec{r}}{\left|\vec{R}-\vec{r}\right|^{3}}\ket{n'\;l'\;m'}.\nonumber
\end{eqnarray}
Energies of the states of the combined atom-molecule system can be obtained by diagonalizing the Hamiltonian Eq.(\ref{eq:hamiltonian}). 
Details of the calculation of matrix elements Eq.(\ref{eq:matrix-elements}) are given in Appendix A.

\section{Numerical results for energies of $KRb-Rb$ and $RbYb-Rb$ systems}

We considered two polar molecules KRb and RbYb of particular interest in the ultracold community. KRb with a permanent electric dipole moment and 
rotational constant 
of $d=0.566$ D \cite{KRb-dip-mom} and $B=1114$ MHz \cite{KRb-rot-const}, respectively, was the first polar molecule produced in the 
ground rovibrational electronic $^{1}\Sigma^{+}$  state at ultracold temperatures \cite{KRb-ground-state} and is the most experimentally 
well-mastered at the moment. RbYb with $d\approx 1$ D \cite{RbYb-dip-mom-rot-const} and 
$B=353$ MHz \cite{RbYb-dip-mom-rot-const} belongs to the family of open-shell molecules with 
$^{2}\Sigma^{+}$ ground electronic state, and is actively studied experimentally \cite{RbYb-experim,RbYb-hyperfine} and theoretically 
\cite{RbYb-theory} towards the goal of producing ground rovibrational state molecules. 
Polar molecules with the $^{2}\Sigma^{+}$ ground state have both an electric and a magnetic dipole moment and are attractive for applications in quantum 
computation \cite{Param-mol-QC} 
and simulation of lattice-spin models \cite{Param-mol-QSim}.
Other candidate molecules with 
subcritical dipoles, to which the readout method is applicable, include RbCs ($d=1.25$ D \cite{RbCs-dip-mom}, $B=490$ MHz \cite{RbCs-rot-const}) and LiNa ($d=0.566$ D, $B=11.3$ GHz \cite{LiNa}) among alkali dimers, and 
a number of alkali metal-alkaline earth diatoms such as NaSr ($d=0.63$ D, $B=1.89$ GHz \cite{Alkali-metal-alkaline-earth}), KSr ($d=1.5$ D, $B=960$ MHz \cite{Alkali-metal-alkaline-earth}), 
RbSr ($d=1.53$ D, $B=540$ MHz \cite{Alkali-metal-alkaline-earth}), and NaCa ($d=1$ D, $B=2.49$ GHz \cite{NaCa}). 
In Rb, the $(n-3)l$ with $l>3$ degenerate manifolds are known to produce considerable mixing of the Rydberg energies, leading to 
formation of large permanent dipole moments in the polyatomic molecules \cite{Seth-Hossein-Others-Science-2011}. To be able to manage the size 
of the Hamiltonian matrix, we do not take account of these degenerate manifolds. In fact, the inclusion of such terms should yield more 
pronounced shifts of the molecular lines and hence better visibility for rotational qubit addressing.

In RbYb, the unpaired electron spin couples to the Rb nuclear spin, resulting 
in hyperfine splitting of the ground electronic state, which is expected to be close to the splitting of $6.835$ GHz between $F=1$ and $F=2$ hyperfine states 
of Rb atom \cite{RbYb-hyperfine}. In the calculations RbYb is assumed to be in the ground electronic potential, corresponding to the lowest in energy $F=1$ hyperfine state. Rotational states of the ground state of RbYb are further split by a spin-rotation interaction 
$\gamma_{SR}\vec{J}\vec{S}$, whose coupling strength can be approximated as $\gamma_{SR}=-2\Delta g_{\perp}B$ \cite{Curl}, 
where $\Delta g_{\perp}=g_{\perp}-g_{e}$ is the deviation of the molecular g tensor component, perpendicular to a molecular axis, from 
the electron's value. The spin-rotation splittings have not been detected for $J=1$ rotational states of the last and second last bound vibrational levels 
of RbYb \cite{Munchow-thesis}. In this experiment the frequency resolution was $\Delta f_{res}\approx 6$ MHz, limiting the spin-rotation constant 
to this value. The rotational constant for such high vibrational states was measured to be $B(\nu=-1) \approx 30$ MHz, while for the ground vibrational state it is predicted 
to be $B(\nu=0)=353$ MHz, setting an upper limit on the spin-rotation constant in the ground vibrational state $\sim \Delta f_{res}B(\nu=0)/B(\nu=-1)\approx 70$ MHz,  
provided $\Delta g_{\perp}$ does not significantly vary with the vibrational number. 

The effect of the 
spin-rotation splitting (not taken into account in the calculations) can be estimated in the following 
way: if $|V_{\rm e^--M}| \ll 2B$, the energy shifts of the atom-molecule system can be approximated using perturbation theory 
as $|\Delta E| \sim |V_{\rm e^--M}|^{2}/2B$. The spin-rotation will modify the energy shifts as 
$|\Delta E| \sim |V_{\rm e^--M}|^{2}/(2B \pm \gamma_{\rm sr}) \approx |V_{\rm e^--M}|^{2}\left(1 \pm \gamma_{\rm sr}/2B\right)/2B$, 
where $\gamma_{\rm sr}/2B \le 0.1$.

\begin{figure*}
\center{
\includegraphics[width=14.cm]{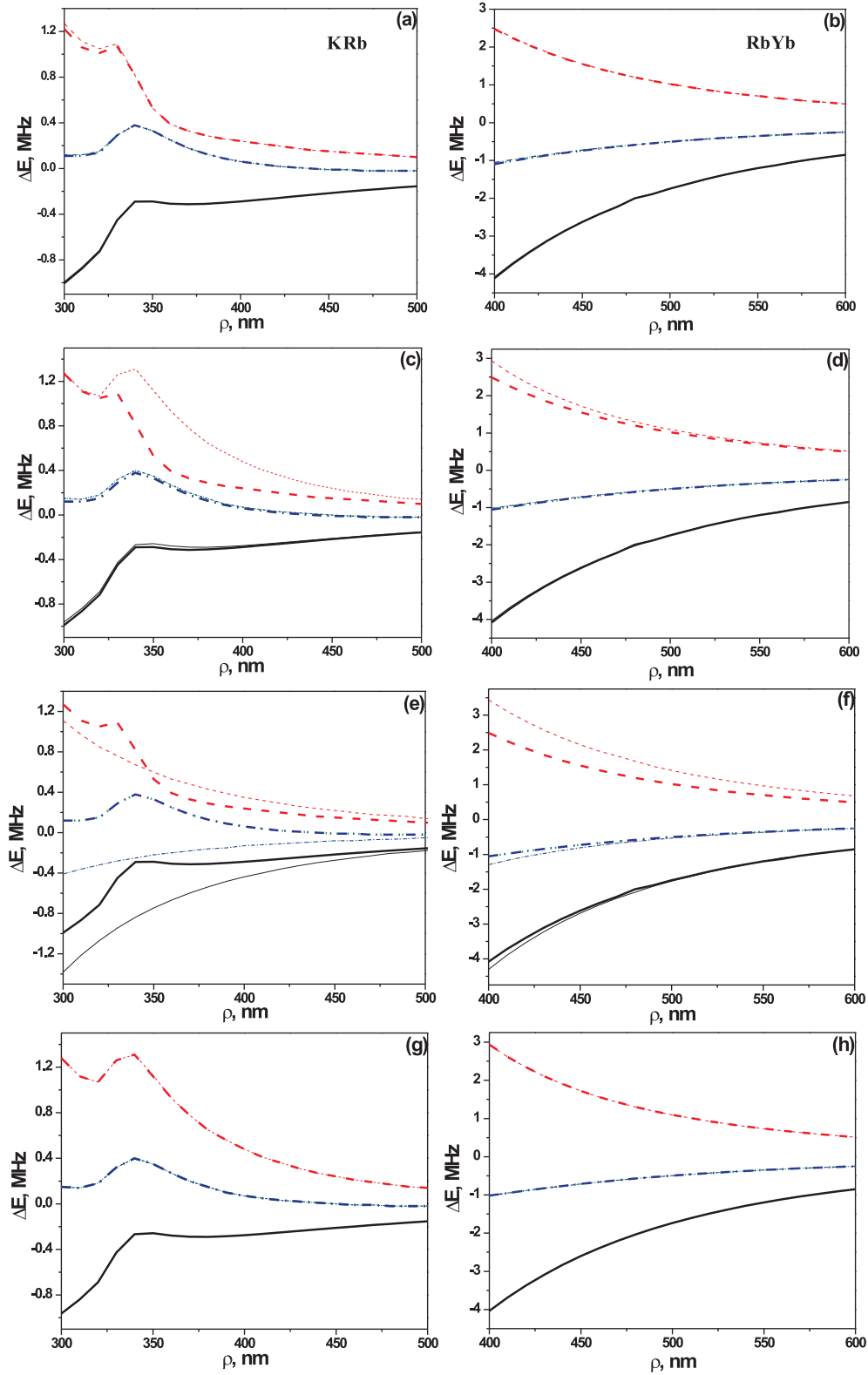}
\caption{\label{fig:shifts-full} (a) and (b) Shifts of the states $\ket{ns}\ket{J,m_{J}}$ of the combined single Rydberg atom-single polar molecule 
system for KRb and RbYb, respectively, 
and Rb, taking into account atomic 60s, 60p, 59p, 59d, 58d and 57f and molecular rotational $J=0,1,2$ states. Shifts of the  $\ket{60s}\ket{J=0,m_{J}=0}$, 
$\ket{60s}\ket{J=1,m_{J}=0}$ and $\ket{60s}\ket{J=1,m_{J}=\pm 1}$ states are shown by thick solid black, thick dashed red, 
thick dotted green and thick dash-dotted blue lines, respectively. Thin lines correspond to calculations in which 57f 
state has not been taken into account;
In (c) the full shifts of (a), shown by thick lines, are compared for KRb to the shifts calculated using a smaller atomic basis set of $60s$, $60p$, $59p$ and $58d$, 
shown by thin lines; (d) the same as in (c) for RbYb; 
In (e) the full shifts of (a), shown by thick lines are compared for KRb to the shifts calculated using the smallest atomic basis set of $60s$, 
shown by thin lines; (f) the same as (e) for RbYb; (g) the 
shifts calculated using the smaller atomic basis set of $60s$, $60p$, $59p$, $58d$ and 
$J=0,1,2$ rotational states (solid lines) are compared to the shifts calculated using $J=0,1,2,3$ rotational states for KRb; 
(h) the same as (g) for RbYb. }
}
\end{figure*}

In Fig.\ref{fig:shifts-full}a and Fig.\ref{fig:shifts-full}b 
the energy shift of the states $\ket{J=0,m_{J}=0}$ and $\ket{J=1,m_{J}=0,\pm 1}$, 
are shown for KRb and RbYb, respectively, interacting with Rb($ns=60s$) state. These states contain admixtures of other states of the 
order of $\le 0.2\%$. In the figures, the rotational splitting $2B$ between the $J=0$ and $J=1$ states has been subtracted. 

In Fig.\ref{fig:shifts-full}b, one observes that for RbYb the splitting between the states $\ket{J=0,m_{J}=0}$ and $\ket{J=1,m_{J}=0}$ 
lies in the range $\sim 6.5-1.2$ MHz for $\rho \sim 400-600$ nm, and the states $\ket{J=0,m_{J}=0}$ and $\ket{J=1,m_{J}=\pm 1}$ are split in the range $\sim 3-0.6$ MHz. Due to a smaller permanent 
dipole moment of KRb and a larger rotational constant the splittings for KRb are 
smaller compared to the splittings for RbYb at the same $\rho$. Splittings of the order $\sim 1$ MHz 
can be achieved for KRb for smaller $\rho$. As  
shown in Fig.\ref{fig:shifts-full}a  
the states $\ket{J=0,m_{J}=0}$ and $\ket{J=1,m_{J}=0}$ are split by $\sim 2.2-0.4$ MHz for $300 \;{\rm nm}<\rho < 500\;{\rm nm}$, and the states 
$\ket{J=0,m_{J}=0}$ and $\ket{J=1,m_{J}=\pm 1}$ 
are split by $\sim 1.1-0.2$ MHz for the same range. The splittings are much larger than the width of Rb($60s$) $\Gamma_{60s} \approx 1.644$ kHz, resulting from spontaneous emission, black-body radiation (BBR) induced 
decay, excitation and 
ionization and the width of the $J=1$ rotational state due to spontaneous emission and interaction with black-body radiation \cite{Rydb-decay-rates}. 

  Next we checked if all the states $np$, $(n-1)p$, $(n-1)d$, $(n-2)d$ and $(n-3)f$ are contributing significantly to the energy shifts. 
Figs.\ref{fig:shifts-full}c and d show the splittings taking into account the full set of 60s, 60p, 59p, 59d, 58d and 57f states (thick lines) and 
and three closest in energy $60p$, $59p$ and $58d$ states (thin lines) for KRb and RbYb, respectively. One can see that for 
RbYb the smaller basis set gives a good agreement with the full one except for the $\ket{J=1,m_{J}=0}$ state, which has an energy difference $\le 16\%$
with the full set. For KRb the agreement between the two sets for the $\ket{J=1,m_{J}=0}$ is worse, the smaller and the full sets giving the same results only at $\rho \approx 300$ nm. 
We also compared the shifts calculated using the full
 set and the smallest possible atomic set including only the $60s$ state. It will give a good 
approximation to the full set if $2B \ll E_{np}-E_{ns},\; E_{ns}-E_{(n-1)p},\; E_{ns}-E_{(n-2)d}$, 
which is satisfied for RbYb and to a less extent for KRb. The results are shown in Fig.\ref{fig:shifts-full}e and f for KRb and RbYb, respectively. 
Again, for RbYb the smallest set has a good agreement with the full one for the $\ket{J=0,m_{J}=0}$ 
state, and the energy difference between the two sets is 
$\le 32\%$ 
for the $\ket{J=1,m_{J}=0}$ state and $\le 20\%$ for the $\ket{J=1,m_{J}=\pm 1}$ state. 
 For KRb 
the shifts for the smallest set reasonably agree with full set shifts only at $450 {\rm nm} < \rho < 500 {\rm nm}$, where the error for 
$\ket{J=0,m_{J}=0}$ is $\le 23\%$, and $\le 38\%$ for 
$\ket{J=1,m_{J}=0}$.

The effect of the 57f state can be seen in Fig.\ref{fig:shifts-full}a and b for KRb and RbYb, respectively. The thick lines show the shifts calculated 
using the full set, and the thin lines correspond to shifts calculated without the 57f state. The effect of the f state is reasonably small, giving 
the energy difference between the two sets $\le 4\%$ for $\ket{J=1,m_{J}=0}$ state for KRb. For RbYb the 
curves with and without 57f completely overlap.

The above calculations were done taking into account $J=0,1,2$ rotational states. 
Finally, we also checked 
if higher rotational states such as $\ket{J=3,m_{J}=0,\pm 1,\pm 2,\pm 3}$ influence the shifts. Fig.\ref{fig:shifts-full}g and h compare the shifts 
for KRb and RbYb, respectively, using the smaller basis set of $60s$, $60p$, $59p$, $58d$ to simplify the calculations and $J=0,1,2$ (thick curves) and 
$J=0,1,2,3$ (thin curves) rotational states. 
One can see that the curves completely overlap, which means that the shifts of the $J=0$ and $J=1$ rotational states are hardly affected by the $J=3$ states. 

\section{N molecules - Rydberg superatom interaction}

In this section we consider interaction of N polar molecules placed in a 1D array with an array or a 
cloud of atoms in a collective symmetric state with a single Rydberg excitation (superatom). 
To illustrate this idea 
we assume the same setup as in the previous section, in which 
there is a ground state atom placed next to each molecule in a linear atomic 1D array, parallel to the molecular 
one. The difference with the 
previous section is that a laser beam, exciting atoms to a Rydberg state, is assumed to interact 
simultaneously with all of them. If the size of the atomic 
array is smaller than the dipole blockade radius \cite{Dipole-blockade}, only one of the atoms will be excited to the Rydberg state, and the array 
wavefunction will be a superposition of states with different single excited  atoms 

\begin{eqnarray}
\label{eq:superatom-state}
\ket{\Psi_{\rm atom}}=\frac{1}{\sqrt{N_{a}}}\sum_{j=1}^{N_{a}}e^{ik_{x\;{\rm Rydb}}X_{j}}\ket{g_{1},...g_{j-1},r_{j},g_{j+1},...g_{N_{a}}},
\end{eqnarray}
where $N_{a}$ is the number of atoms, $\ket{g_{j}}$ and $\ket{r_{j}}$ denote j$^{\rm th}$ atom in the 
ground or Rydberg state and $X_{j}$ its position along the array, $k_{x\;{\rm Rydb}}$ is the x component of the wavevector of 
the exciting laser 
field. 
Here, we take into account a single Rydberg state $\ket{r}=\ket{60s}$ for each atom in 
diagonalizing the interaction Hamiltonian $V_{\rm e^--M}$. Then the interaction of i$^{\rm th}$ molecule with the 
atomic array takes the form:
\begin{eqnarray}
\label{eq:matrix-element-many-body}
\lefteqn{\bra{\Psi_{\rm atom}}V_{\rm e^--M,\; i}\ket{\Psi_{\rm atom}} \;=} \\
& = & \frac{1}{N_{a}}\sum_{j=1}^{N_{a}}\bra{r_{j}}V_{\rm e^--M,\;i}\ket{r_{j}} \;=\;
\frac{1}{N_{a}}\sum_{j=1}^{N_{a}}V_{ji},\nonumber
\end{eqnarray}
where $V_{ji}=\bra{r_{j}}V_{\rm e^--M,\; i}\ket{r_{j}}=\bra{r_{j}}\frac{e\vec{d}_{i} \cdot \vec{R}_{ji}}{R_{ji}^{3}}-\frac{e\vec{d}_{i} \cdot (\vec{R}_{ji}-\vec{r})}{|\vec{R}_{ji}-\vec{r}|^{3}}\ket{r_{j}}$ 
and $\vec{R}_{ji}$ is the vector connecting j$^{\rm th}$ atom to i$^{\rm th}$ molecule. In fact, one can see from Eq.(\ref{eq:matrix-element-many-body})  
that interatomic coherences do not play a role and the same result can be obtained with a mixed atomic state described by a density matrix
\[
\rho=\frac{1}{N_{a}}\sum_{j=1}^{N_{a}} \ket{\phi_j}\bra{\phi_j}
\]
where $\ket{\phi_j} = \ket{g_1\ldots g_{j-1}r_jg_{j+1}\ldots g_N}$.

The basis states of the combined atomic-molecular system are $\ket{\Psi_{\rm atom}}\ket{\Psi_{\rm mol}}$, where $\ket{\Psi_{\rm mol}}=\ket{a_{1}\ldots a_{N}}$ and $\ket{a_{i}}=\ket{J,m_{J}}_{i}$. Matrix elements of the Hamiltonian of the system 
\begin{equation}
\label{eq:many-body-hamiltonian}
H=H_{\rm a}+H_{\rm m}+V_{\rm e^--M}
\end{equation}
with $H_{\rm a}=\sum_{j=1}^{N_{a}}E_{r_{j}}\ket{r_{j}}\bra{r_{j}}$, $H_{\rm m}=\sum_{i=1}^{N}B{\bf J}^{2}_{i}$ and 
$V_{\rm e^--M}=\sum_{j=1}^{N_{a}}\sum_{i=1}^{N}\frac{e\vec{d}_{i} \cdot \vec{R}_{ji}}{R_{ji}^{3}}-\frac{e\vec{d}_{i} \cdot (\vec{R}_{ji}-\vec{r})}{|\vec{R}_{ji}-\vec{r}|^{3}}$,  
have the form 
\begin{eqnarray}
\label{eq:Ham-many-body}
\lefteqn{\bra{\Psi_{\rm atom}}\bra{\Psi_{\rm mol}}H_{a}\ket{\Phi_{\rm mol}}\ket{\Psi_{\rm atom}}\;=} \\
&=&-\frac{1}{2(ns-\mu_{s})^{2}}\Pi_{i=1}^{N}\delta_{a_{i},a_{i}'}, \nonumber \\[.2cm]
\lefteqn{\bra{\Psi_{\rm atom}}\bra{\Psi_{\rm mol}} H_{\rm m} \ket{\Phi_{\rm mol}}\ket{\Psi_{\rm atom}}\;=} \nonumber \\
&=&\left(\sum_{i=1}^{N}BJ_{i}(J_{i}+1)\right)\Pi_{i=1}^{N}\delta_{a_{i},a_{i}'}, \nonumber \\[.2cm]
\lefteqn{\bra{\Psi_{\rm atom}}\bra{\Psi_{\rm mol}} V_{\rm e^--M} \ket{\Phi_{\rm mol}}\ket{\Psi_{\rm atom}}\;=} \nonumber \\
&=&\left(\frac{1}{N_{a}}\sum_{j=1}^{N_{a}}\bra{a_{i}}V_{ji}\ket{a_{i}'}\right)\Pi_{k=1,k\neq i}^{N}\delta_{a_{k},a_{k}'}\delta_{J_{i},J_{i}'\pm 1}, \nonumber
\end{eqnarray} 
for $i=1\ldots N$, 
 where $\ket{\Phi_{\rm mol}}=\ket{a_{1}',a_{2}',...a_{i}',...,a_{N}'}$.

For a simplified analysis we also take into account only $\ket{J=0,m_{J}=0}$ and 
$\ket{J=1,m_{J}=0}$ rotational states in diagonalizing the Hamiltonian. 
Using only $J=0$ and $J=1$ states makes 
calculations of the matrix elements $\bra{a_{i}}V_{ji}\ket{a_{i}'}$ particularly simple. For the atom located next to the molecules at $\Delta x=0$ 
the matrix 
elements are calculated as described in Appendix A. For an atom separated 
from the molecule by a lattice spacing(s) the calculation of the matrix elements can be reduced to the case of $\Delta x=0$ by a coordinate system 
rotation as shown in Appendix B.

If the interaction strength is much smaller than the rotational splitting between the $J=0$ and $J=1$ states $|V_{\rm e^--M}| \ll E_{\rm rot}$, where $E_{\rm rot}=2B$, and 
only the $ns$ state is taken into account, the shifted energies of the collective rotational states can be 
calculated using second-order perturbation theory. All unperturbed states $(k\uparrow,(N-k)\downarrow)$ with $k$ spins up and $N-k$ spins down 
have the same energy $E_{(k\uparrow,(N-k)\downarrow)}=kE_{\rm rot}$, and groups of states differing by one flipped spin are separated by the rotational 
splitting $E_{\rm rot}$. The interaction weakly couples states in neighboring spin groups resulting in shifts of their energy. 
Let us consider first the $\ket{ns}\ket{\downarrow,\downarrow,...\downarrow}$ 
state for $N$ molecules and $N_{a}$ atoms. The energy shift of this state will be given by:
\begin{eqnarray*}
\lefteqn{\Delta E_{N\downarrow} \;=} \\
&=&-\sum_{i=1}^{N}\frac{|\bra{\Psi_{\rm atom}}\bra{\downarrow,.,\downarrow}V_{\rm e^--M}\ket{\downarrow,.,\uparrow_{i},.,\downarrow}\ket{\Psi_{\rm atom}}|^{2}}{E_{\rm rot}}\;=  \\
&=&-\sum_{i=1}^{N}\frac{|\bra{\downarrow_{i}}\sum_{j=1}^{N_{a}}\bra{ns_{j}}V_{\rm e^--M}\ket{ns_{j}}/N_{a}\ket{\uparrow_{i}}|^{2}}{E_{\rm rot}} \;= \\
&=&-\sum_{i=1}^{N}\frac{|\sum_{j=1}^{N_{a}}V_{\rm e^--M}^{ji}|^{2}}{E_{\rm rot}N_{a}^{2}}
\end{eqnarray*}
Assuming for simplicity that i$^{\rm th}$ molecule most strongly interacts with its nearest atom with $j=i$ and the matrix elements $V_{\rm e^--M}^{j=i}$ 
are the same for all $i$, 
the shift can be approximated as 
\begin{eqnarray}
\Delta E_{N\downarrow} \approx -\frac{\sum_{i=1}^{N}|V_{\rm e^--M}^{j=i}|^{2}}{E_{\rm rot} N_{a}^{2}} \approx -\frac{N|V_{\rm e^--M}^{j=i}|^{2}}{E_{\rm rot} N_{a}^{2}}, \nonumber
\end{eqnarray}
which gives the dependence $\Delta E_{N\downarrow} \sim 1/N$ for $N_{a}\sim N$. For states with a single i$^{\rm th}$ spin up and $i'=1,.,i-1,i+1,.,N$ spins down, the perturbation theory 
gives the energy shift
\begin{widetext}
\begin{eqnarray}
\Delta E_{(1\uparrow,(N-1)\downarrow)} & = & -\sum_{i'=1,i'\ne i}^{N}\frac{|\bra{\Psi_{\rm atom}}\bra{\downarrow,.,\uparrow_{i},.,\downarrow}V_{\rm e^--M}\ket{\downarrow,.,\uparrow_{i},.,\uparrow_{i'},.,\downarrow}\ket{\Psi_{\rm atom}}|^{2}}{E_{\rm rot}}+ \nonumber \\
&&+\frac{|\bra{\Psi_{\rm atom}}\bra{\downarrow,.,\uparrow_{i},.,\downarrow}V_{\rm e^--M}\ket{\downarrow,.,\downarrow}\ket{\Psi_{\rm atom}}|^{2}}{E_{\rm rot}}= \nonumber \\
&&=-\frac{1}{N_{a}^{2}}\sum_{i'=1,i'\ne i}^{N}\frac{|\sum_{j=1}^{N_{a}}V_{\rm e^--M}^{ji'}|^{2}}{2B}+\frac{1}{N_{a}^{2}}\frac{|\sum_{j=1}^{N_{a}}V_{\rm e^--M}^{ji}|^{2}}{E_{\rm rot}}\approx
-\frac{(N-1)}{N_{a}^{2}}\frac{|V_{\rm e^--M}^{j=i'}|^{2}}{E_{\rm rot}}+\frac{1}{N_{a}^{2}}\frac{|V_{\rm e^--M}^{j=i}|^{2}}{E_{\rm rot}}= \nonumber \\
&&=-\frac{(N-2)}{N_{a}^{2}}\frac{|V_{\rm e^--M}^{j=i}|^{2}}{E_{\rm rot}} \nonumber
\end{eqnarray}
\end{widetext}
which shows that the splitting between the $N\downarrow$ and $(1\uparrow,(N-1)\downarrow)$ states $\sim 2|V_{\rm e^--M}|^{2}/E_{\rm rot} N_{a}^{2}\sim 1/N^{2}$. In the general case of 
$(k\uparrow,(N-k)\downarrow)$ states the shift will be given by
\begin{widetext}
\begin{eqnarray}
\label{eq:coll-shifts-pert}
\Delta E_{(k\uparrow,(N-k)\downarrow)} & = & -\frac{1}{N_{a}^{2}}\sum_{i'=1,i'\in (N-k)\downarrow}^{N}\frac{|\sum_{j=1}^{N_{a}}V_{\rm e^--M}^{ji'}|^{2}}{E_{\rm rot}}+
\frac{1}{N_{a}^{2}}\sum_{i=1,i\in k\uparrow}^{N}\frac{|\sum_{j=1}^{N_{a}}V_{\rm e^--M}^{ji}|^{2}}{E_{\rm rot}} \approx  \\
&&\approx -\frac{(N-k)}{N_{a}^{2}}\frac{|V_{\rm e^--M}^{j=i'}|^{2}}{E_{\rm rot}}+\frac{k}{N_{a}^{2}}\frac{|V_{\rm e^--M}^{j=i}|^{2}}{E_{\rm rot}} 
\approx -\frac{(N-2k)}{N_{a}^{2}}\frac{|V_{\rm e^--M}^{j=i}|^{2}}{E_{\rm rot}},  \nonumber
\end{eqnarray}
\end{widetext}
In the case when only the $ns$ atomic state is taken into account there will be no terms exchanging spins within the same group such as 
e.g. $\ket{\downarrow, .,\uparrow_{i},.,\downarrow_{i'},.,\downarrow} \leftrightarrow \ket{\downarrow, .,\downarrow_{i},.,\uparrow_{i'},.,\downarrow}$ 
in the $(1\uparrow,(N-1)\downarrow)$ manifold. The spin-exchange terms will be absent because the second-order perturbation theory connects these states via a single state 
in the upper and a single state in the lower  
neighboring spin groups, which cancel each other due to the equal splitting between neighbouring groups. There will be spin-exchange terms within the groups due to a direct dipole-dipole interaction between molecules, 
which has not been taken into account in the Hamiltonian (\ref{eq:Ham-many-body}). The direct dipole-dipole interaction allows spin-exchange processes 
within the same group leading to splittings $\sim d^{2}/L^{3} \sim 1$ kHz for $d \sim 1$ D and a lattice period $L \sim 500$ nm, which is of the order of the 
width of the 60s state.     

\begin{figure}
\center{
\includegraphics[width=11.cm]{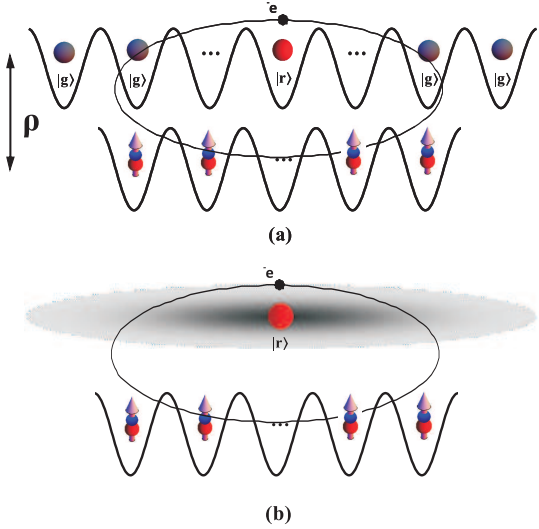}
\caption{\label{fig:many-body-setup} (a) Schematic of the $N$ molecular array interacting with an array of $N_{a}=N+2$ atoms. The first and last atoms are 
added to mitigate the effects of the boundaries. When the atomic system is excited to the 60s state the wavefunction becomes 
$\ket{\Psi_{\rm atom}}=\frac{1}{\sqrt{N_{a}}}\sum_{j=1}^{N_{a}}\ket{g_{1},...r_{j},..,g_{N_{a}}}$, and each molecule equally interacts with all 
(most strongly with three nearest) Rydberg atoms; (b) an array of molecules interacting with a cloud of atoms in the superatom state, placed in an 
elongated dipole trap}
}
\end{figure}

In the following we numerically consider a system of $N=3$ and 5 KRb and RbYb molecules interacting with 
a Rb superatom. Due to the small size of the 
considered molecular arrays, if one uses an atomic array of the same size, effects of the boundaries will be sizable, because the molecules in the center 
will strongly interact with all three nearest atoms, while the molecules at the boundaries will interact strongly with only two atoms. This is not going to be the 
case in a sufficiently long array, in which all molecules (again, except for two boundary ones) will have equal interaction conditions.  
To mitigate the effects of the boundaries we therefore consider an array of $N_{a}=N+2$ atoms, arranged 
in a way shown in Fig.\ref{fig:many-body-setup}a such that there is an additional atom at each side of the molecular array. In this arrangement 
all molecules will interact strongly with three nearest atoms, i.e. separated by at most one 
lattice period $L=500$ nm. Atoms separated from a molecule by two or more lattice periods do not contribute significantly for the $60s$ 
Rydberg state. 

\begin{figure*}
\center{
\includegraphics[width=17.cm]{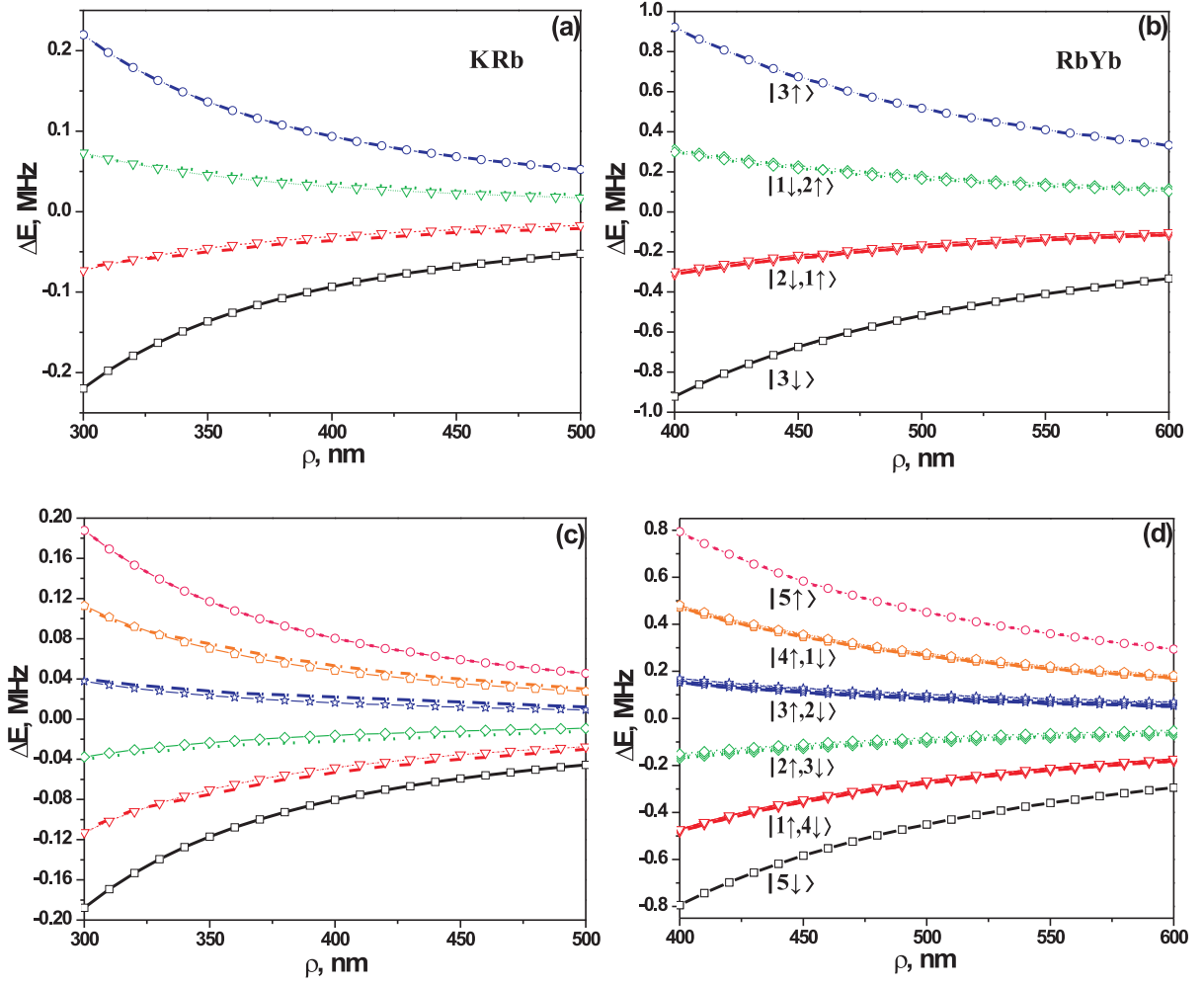}
\caption{\label{fig:many-body-shifts} (a) and (b) Shifts of the states of the combined system of 
N=3 polar molecules and $N_{a}=N+2$ atoms for KRb and 
RbYb, respectively, and Rb(60s) calculated by diagonalizing Hamiltonian (\ref{eq:Ham-many-body}) (lines) and using the perturbation theory 
expression (\ref{eq:coll-shifts-pert}) (open symbols). 
The atomic system 
is in the state $\ket{\Psi_{\rm atom}}=\frac{1}{\sqrt{N_{a}}}\sum_{j=1}^{N_{a}}\ket{g_{1},...r_{j},..,g_{N_{1}}}$, and only $60s$, 
$\ket{\downarrow}=\ket{J=0,m_{J}=0}$ and $\ket{\uparrow}=\ket{J=1,m_{j}=0}$ 
 states have been used in diagonalizating (\ref{eq:Ham-many-body}). 
Solid black line (open squares) correspond to $\ket{3\downarrow}=\ket{\downarrow,\downarrow,\downarrow}$ state, dashed red lines (open triangles) 
correspond to single spin up states $\ket{2\downarrow,1\uparrow}$, dotted green lines (open diamonds) correspond to two spins up 
$\ket{1\downarrow,2\uparrow}$ 
states and a dashed-dotted blue line (open circles) correspond to three spins up $\ket{3\uparrow}$ state; (c) and (d) the same as in (a) and (b) 
but for $N=5$ molecules: solid black line (open squares) correspond to $\ket{5\downarrow}$ state, dashed red lines (open triangles) correspond to 
$\ket{1\uparrow,4\downarrow}$ states, dotted green lines (open diamonds) correspond to $\ket{2\uparrow,3\downarrow}$ states, dash-dotted blue lines 
(open stars) correspond to $\ket{3\uparrow,2\downarrow}$ states, dash-dot-dotted orange lines (open pentagons) correspond to $\ket{4\uparrow,1\downarrow}$ 
states, and short dash pink line (open circles) correspond to the $\ket{5\uparrow}$ state.}
}
\end{figure*}

Energy shifts of the states of a combined atomic-molecular system from the unperturbed values $E_{(k\uparrow,(N-k)\downarrow)}$, corresponding to different collective rotational states, are 
given in Fig.\ref{fig:many-body-shifts} for KRb and RbYb in the left and right columns, respectively. Since only the 60s Rydberg state and $J=0$ and $J=1$ rotational states 
have been used in calculation of matrix elements of the Hamiltonian Eq.(\ref{eq:many-body-hamiltonian}), the 
magnitude of the shifts is only qualitatively correct. In order to get quantitatively correct shifts the closest in energy p, d and f
Rydberg states as well as $J=2$ rotational states have to be taken into account, which is beyond the scope of our work. Fig.\ref{fig:many-body-shifts} 
shows the shifts calculated by (i) diagonalization of the Hamiltonian Eq.(\ref{eq:many-body-hamiltonian}) and (ii) using perturbation theory expression 
 (\ref{eq:coll-shifts-pert}) with $V_{\rm e^--M}^{ji}$ calculated numerically. The results of the two calculations agree very well.

From Fig.\ref{fig:many-body-shifts} one can see that the states with the same number of spins up and down 
such as $\ket{\uparrow,\downarrow,\downarrow}$, $\ket{\downarrow,\uparrow,\downarrow}$ and 
$\ket{\downarrow,\downarrow,\uparrow}$ group, as expected, so only states with at least one spin flipped significantly 
differ in energy and can be discerned. For $N=3$ 
the spin groups 
split from each other by $\sim 150-50$ kHz for $300 \; {\rm nm}\; <\rho < \; 500 \;{\rm nm}$ for KRb and 
by $\sim 600-200$ kHz for 
$400 \; {\rm nm}\; <\rho < \; 600 \;{\rm nm}$ for RbYb. For an array of $N=5$ 
molecules, shown in Fig.\ref{fig:many-body-shifts}c and d, the energy splittings between spin groups become smaller:  
$\sim 70-10$ kHz for KRb, and $\sim 300-100$ kHz for RbYb in the same ranges of $\rho$. 
The reduction of the splittings with increasing $N$ comes from the 
simultaneous increase in $N_{a}$ as expected from Eq.(\ref{eq:coll-shifts-pert}).
It suggests that for larger $N$ the splittings will get even smaller and eventually become comparable to the width of the 
Rydberg state. The states will no longer be discernable and there will be a continuous band of collective molecular states.

\begin{figure*}
\center{
\includegraphics[width=17.cm]{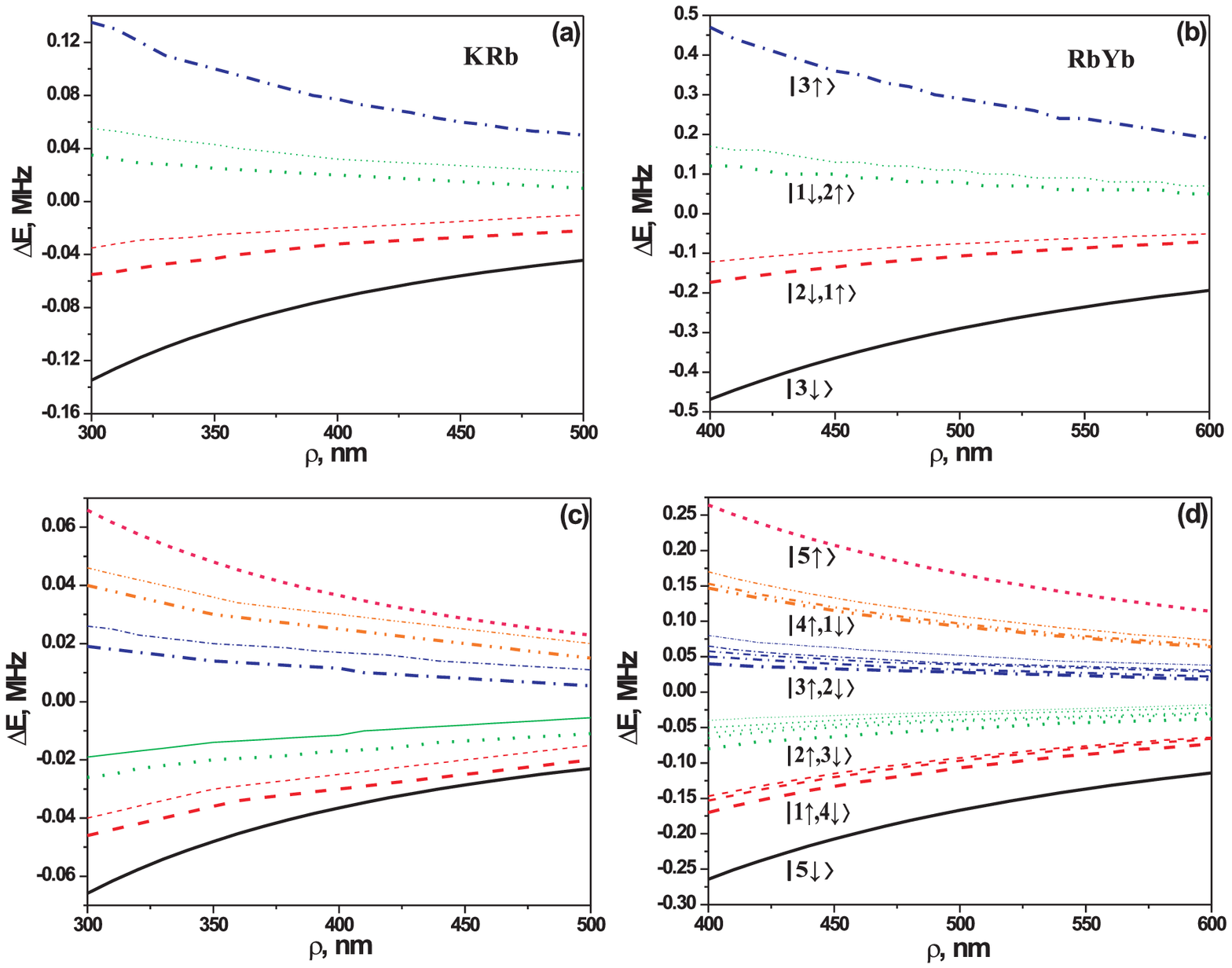}
\caption{\label{fig:many-body-shifts-aver} (a) and (b) Shifts of the states of N=3 polar molecules of KRb and 
RbYb, respectively, and a Rb superatom in 60s, calculated by diagonalizing the Hamiltonian Eq.(\ref{eq:Ham-many-body}). The atoms are placed in a 1D Gaussian trap along molecular array with the center of the trap corresponding 
to the center of the array. The probability to find a single Rydberg atom scales as $p(x)=\exp{(-x^{2}/a^{2})}/\sqrt{\pi}a$ along the trap. The trap widths 
are $a=1.3$ $\mu$m and $a=1.7$ $\mu$m for KRb and RbYb, respectively. 
(c) and (d) shifts of collective states for $N=5$ molecules of KRb and RbYb, respectively. The trap widths are $a=2.5$ $\mu$m for KRb and $a=3$ $\mu$m for RbYb. 
Collective rotational states are denoted similarly to Fig.\ref{fig:many-body-shifts} (detailed description is given in \cite{Many-body-description}).}
}
\end{figure*}

In experiments it can be difficult to have two parallel optical lattices, one filled with molecules and another with atoms. The setup will be simplified 
if the atoms are placed in a sigar-shaped dipole trap with a long axis parallel to 
the molecular array as shown in Fig.\ref{fig:many-body-setup}b. We 
modelled the interaction of a molecular array with a Rydberg superatom placed in such a trap by assuming 
that the excited atom has a Gaussian 1D probability distribution 
$p(x)=\exp(-x^{2}/a^{2})/a\sqrt{\pi}$ along the longest trap axis with $x=0$ corresponding to the center 
of the molecular array. In this case the summation 
$\sum_{j=1}^{N_{a}}V_{ji}/N_{a}$ 
over atom's $j$ position in Eqs.(\ref{eq:matrix-element-many-body}) and (\ref{eq:Ham-many-body}) is replaced by an integral $\int V_{i}(x)p(x)dx$.
Fig.\ref{fig:many-body-shifts-aver}a(c) and b(d) show the shifts of the collective states for $N=3(5)$ for KRb and RbYb, respectively, obtained 
by diagonalization of (\ref{eq:Ham-many-body}). The size of the 
atomic distribution $a$ was chosen to have the largest splittings between spin states differing by one flipped spin. Compared to the case of atoms in a lattice the 
shifts between spin groups become about twice smaller for $N=3$ and three times smaller for $N=5$ for both 
KRb and RbYb. The splittings 
within groups appear, and for $N=5$ become comparable to the splittings between the groups. The 
reason for splittings within the groups is in the unequal interaction conditions for molecules in the center and at the boundaries. Due to the 
decrease of the atom's probability from the center to the edges of the trap the corresponding molecules will experience weaker interaction. 
The interaction conditions can be made more homogeneous if the trap size in the longitudinal direction is much larger than the molecular array's, but 
in this case the probability to find the atom in the range $(X_{i}-L/2,X_{i}+L/2)$ around i$^{\rm th}$ molecule position $X_{i}$ will be smaller 
than the corresponding probability $1/N_{a}$ of atoms in a lattice.

The $\sim 1/N$ and $\sim 1/N^{2}$ scalings 
of the spin group energy shifts and splittings can be avoided if the molecules are placed in a ring 1D array \cite{Ring-lattice} instead of a linear one. 
In this case if the superatom (or a single Rydberg atom) is  placed 
at the center of the array and its size is much smaller than the radius of the array all interaction matrix elements $V_{\rm e^--M}^{ji}$ will be 
equal for a symmetric $ns$ state $V_{\rm e^--M}^{ji}={\tilde V}_{\rm e^--M}$ and the shifts (\ref{eq:coll-shifts-pert}) will become:
\begin{eqnarray*}
\Delta E_{(k\uparrow,(N-k)\downarrow)}&=&-(N-k)\frac{|{\tilde V}_{\rm e^--M}|^{2}}{E_{\rm rot}}+k\frac{|{\tilde V}_{\rm e^--M}|^{2}}{E_{\rm rot}} \approx\\
&\approx& -(N-2k)\frac{|{\tilde V}_{\rm e^--M}|^{2}}{E_{\rm rot}},
\end{eqnarray*}
which shows that in this case the shifts and splittings scale as $\Delta E \sim N|{\tilde V}_{\rm e^--M}|^{2}/E_{\rm rot}$ and 
$\sim 2|{\tilde V}_{\rm e^--M}|^{2}/E_{\rm rot}$ with the number of the molecules,  
so the splittings are limited by the interaction strengths $|{\tilde V}_{\rm e^--M}|^{2}$, falling with the atom-molecule distance, equal to the 
radius of the array, as 
$1/R_{ji}^{4}$.

Selective excitation to the Rydberg state for a particular spin group will 
require the Rabi frequency $\Omega$ of an exciting optical pulse be smaller than the splittings. With the splittings between spin groups 
$\sim$ hundreds kHz one can use $\Omega=10$ kHz. In this case the blockade radius for the $60s$ state will be 
of the order of $R_{b}=\left(C_{6}/\hbar \Omega\right)^{1/6} \sim 2.2$ $\mu$m, where $C_{6}$ is taken from \cite{Robin-Rydb}. For the linear 
array period $L=500$ nm the blockade can be realized for $N_{a}\le 9$ atoms in $60s$, i.e. higher $n$ are required for larger arrays. For example, for $n=100$ 
and the same Rabi frequency $\Omega=10$ kHz the blockade radius will be $\approx 44$ $\mu$m, which will allow to use a linear array of 
$\sim 200$ atoms and molecules. In a ring array any number of atoms can be used provided the atomic trap size is smaller than the blockade radius.

\section{Measurement of rotational state population}

Let us first discuss the measurement of rotational states population of a single molecule interacting with a single Rydberg atom. 
In this case we assume that each molecule in an array or optical lattice has its own readout atom, which can be addressed individually 
using tightly focussed laser beams and excited to the Rydberg state without affecting neighboring atoms. 
Individual addressing of atoms in  an optical lattice with a lattice period $\lambda/2 \sim 500$ nm has been demonstrated recently in 
\cite{Atom-microscope}. Alternatively, low cross-talk addressing of atoms can be realized if, before the 
readout, the molecular and atomic lattices period is increased to values $\sim $ several $\mu$m using tunable-period 1D or 2D optical lattices 
\cite{Accordeon-lattice}.

\begin{figure*}
\center{
\includegraphics[width=17.cm]{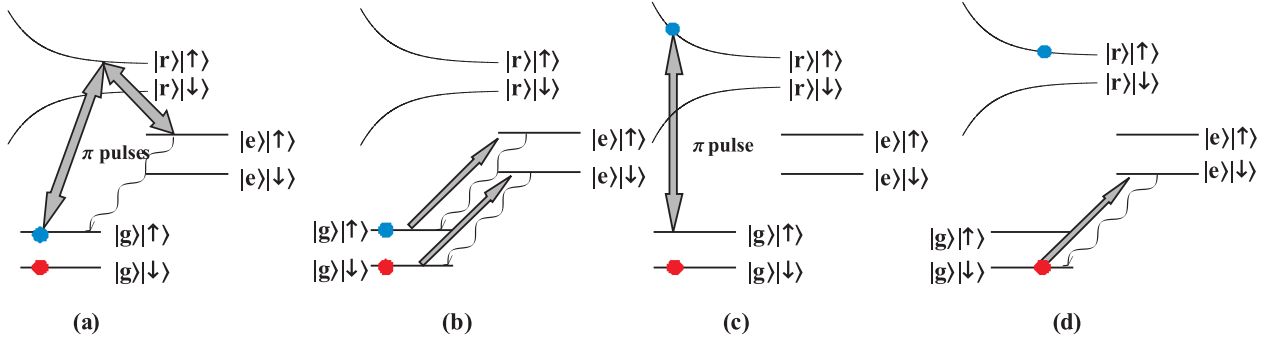}
\caption{\label{fig:fluorescence-readout} Readout of populations of molecular rotational states $\ket{\downarrow}=\ket{J=0,m_{J}=0}$ and 
$\ket{\uparrow}=\ket{J=1,m_{J}=0,\pm 1}$. (a) Population of the $\ket{\uparrow}$ state can be measured by selectively exciting 
the atom-molecule system to the $\ket{r}\ket{\uparrow}$ state by a $\pi$ pulse, followed by transfer by a second $\pi$ pulse to some $\ket{e}\ket{\uparrow}$ state, rapidly decaying 
to the $\ket{g}\ket{\uparrow}$ state. The transfer $\ket{g}\ket{\uparrow} \rightarrow \ket{r}\ket{\uparrow} \rightarrow \ket{e}\ket{\uparrow}$ can be also done using 
STIRAP. The population of the $\ket{g}\ket{\downarrow}$ can be read out in a similar way; (b) While the atom is in the ground state and does not interact with the molecule 
atomic fluorescence intensity can be measured using excitation and deexcitation on a cycling transition $\ket{g}\leftrightarrow \ket{e}$; 
(c) The combined system can be 
conditionally transferred from the $\ket{g}\ket{\uparrow}$ to the $\ket{r}\ket{\uparrow}$ state vis a $\pi$ pulse or an ARP pulse; (d) Atomic fluorescence intensity 
on the $\ket{g} \leftrightarrow \ket{e}$ transition is measured again. The difference between the fluorescence intensities before and after the Rydberg excitation allows to 
obtain populations of the rotational states. }
}
\end{figure*}

The populations of the 
ground and first excited rotational states can be read out in the following way (see Fig.\ref{fig:fluorescence-readout}a). First, the atom is prepared in the 
ground state, e.g.  $\ket{g}=\ket{F=2,m_{F}=2}$ for Rb, such that the state of the combined system is $(\alpha \ket{\downarrow}+\beta \ket{\uparrow})\ket{g}$, where $|\alpha|^{2}$ and $|\beta|^{2}$ 
are the molecular qubit states populations to be read out. The combined system next is transferred selectively 
from e.g. the $\ket{g}\ket{\uparrow}$ to the $\ket{r}\ket{\uparrow}$, where $\ket{r}$ is the atomic Rydberg state, with a $\pi$ pulse, followed by another $\pi$ pulse transferring the system from 
the $\ket{r}\ket{\uparrow}$ to some axiliarly $\ket{e}\ket{\uparrow}$ state with a short lifetime, which rapidly decays back to 
$\ket{g}\ket{\uparrow}$ \cite{STIRAP-excitation}. 
Repeating the excitation-fluorescence cycle one can detect the population $|\beta|^{2}$ of the $\ket{\uparrow}$ rotational state. 
Another way to read out populations of molecular qubits is shown in Fig.\ref{fig:fluorescence-readout}b,c and d.
 While the atom is in the ground state its fluorescence is measured using the cycling transition $\ket{g}\rightarrow \ket{e}$, which will include contributions from both rotational states with probabilities 
$|\alpha|^{2}$ and $|\beta|^{2}$ (Fig.\ref{fig:fluorescence-readout}b). Next a $\pi$ optical pulse is applied 
to excite the system conditionally only if the 
molecule is in the $\ket{\uparrow}$ state, producing an entangled state of the combined system  
$\alpha \ket{\downarrow}\ket{g}+\beta \ket{\uparrow}\ket{r}$, and again the atomic fluorescence is measured using the cycling transition (Fig.\ref{fig:fluorescence-readout}c). This time the part of the 
atom in the Rydberg state will not fluoresce and only the $\ket{g}\ket{\downarrow}$ part will contribute with the probability $|\alpha|^{2}$ (Fig.\ref{fig:fluorescence-readout}d). 
By measuring the ratio of fluoresence intensities in the two cases $R=|\alpha|^{2}/(|\alpha|^{2}+|\beta|^{2})$ one can calculate the populations of the rotational states 
$|\alpha|^{2}=R$ and $|\beta|^{2}=1-R$. The Rydberg state has a finite lifetime and can decay during the readout steps of Fig.\ref{fig:fluorescence-readout}c,d. 
The decay can be avoided if  the 
atom is transferred to a stable ground state $\ket{g'}$ by a $\pi$ pulse: 
$\ket{g}\ket{\uparrow} \rightarrow \ket{r}\ket{\uparrow} \rightarrow \ket{g'}\ket{\uparrow}$ such that $\ket{g'}$ is not affected by the excitation-fluorescence 
cycles $\ket{g} \leftrightarrow \ket{e}$. Additionally, it allows one to entangle two atomic ground states and two rotational molecular states as
$(\alpha \ket{\downarrow}+\beta \ket{\uparrow})\ket{g} \rightarrow \alpha \ket{g}\ket{\downarrow}+\beta \ket{g'}\ket{\uparrow}$.

The main part of the measurement described in 
Fig.\ref{fig:fluorescence-readout} is a CNOT gate applied to the atom-molecule system. 
The measurement based on the CNOT gate is of a quantum non-demolishion type, in which the 
measurement is done on the ancilla system after it has interacted with the primary system in such a way that the primary system is not 
destroyed and it's projected states are not disturbed by the measurement. In \cite{QND-theory} requirements for a QND measurement 
 on a primary qubit by an ancilla qubit have been derived in terms of fidelities of measurement 
$F_{\rm M}=\sqrt{\sum_{i}p_{i}^{\rm M}p_{i}^{\rm in}}$, QND fidelity $F_{\rm QND}=\sqrt{\sum_{i}p_{i}^{\rm in}p_{i}^{\rm out}}$ and the quantum state 
preparation fidelity $F_{\rm QSP}=\sum_{i}p_{i}^{\rm M}p_{\ket{i}|i}^{\rm out}$, where $p_{i}^{\rm in}$, $p_{i}^{\rm M}$ and $p_{i}^{\rm out}$ are the 
probability distributions of the input, measured and output states in the basis of the eigenstates $\ket{i}$ of the measurement; 
$p_{\ket{i}|i}^{\rm out}$ is the conditional probability of finding the output state to be $\ket{i}$ if the measurement gave the eigenvalue $i$. 
The CNOT gate gives $F_{\rm M}=F_{\rm QND}=F_{\rm QSP}=1$, i.e. it represents an ideal QND measurement. The QND nature of the measurement can be seen from the 
form of the actual atom-molecule interaction 
\begin{eqnarray*}
\lefteqn{\ket{ns}\bra{ns}\left(\Delta E_{\uparrow}\ket{\uparrow}\bra{\uparrow}+\Delta E_{\downarrow}\ket{\downarrow}\bra{\downarrow}\right) \;= } \\
&=&\left(\hat{S}^{\rm at}_{z}+\frac{1}{2}\right)\left(\Delta E_{\uparrow}(\hat{S}^{\rm mol}_{z}+\frac{1}{2})+\Delta E_{\downarrow}(\frac{1}{2}-\hat{S}^{\rm mol}_{z})\right)\\
&=& (\Delta E_{\uparrow}-\Delta E_{\downarrow})\hat{S}^{\rm at}_{z}\hat{S}^{\rm mol}_{z}+\hat{S}^{\rm at}_{z}\frac{\Delta E_{\downarrow}+\Delta E_{\uparrow}}{2}+ \\
&&+\hat{S}^{\rm mol}_{z}\frac{\Delta E_{\downarrow}-\Delta E_{\uparrow}}{2}, 
\end{eqnarray*}
which commutes with the measured observable $\hat{S}^{\rm mol}_{z}$. 
Here $\hat{S}_{z}^{\rm at}=(\ket{ns}\bra{ns}-\ket{g}\bra{g})/2$ and $\hat{S}_{z}^{\rm mol}=(\ket{\uparrow}\bra{\uparrow}-\ket{\downarrow}\bra{\downarrow})/2$.

The QND measurement based on the CNOT gate has been used previously in systems of two ions \cite{QND-ions}, electron-nuclear spins of N-V center 
\cite{QND-N-V} and was also theoretically discussed for a system of two neutral atoms of different species, interacting in Rydberg states \cite{Beterov-Saffman-QLS}. 

\begin{figure}
\center{
\includegraphics[width=9.cm]{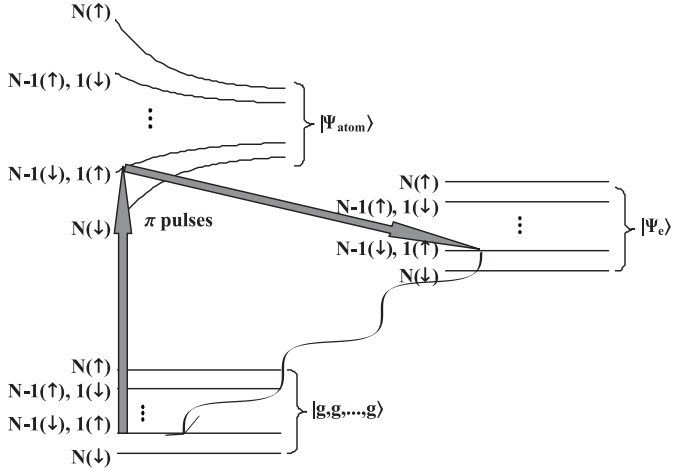}
\caption{\label{fig:many-body-fluorescence} Schematic of measurement of populations of collective states with a certain number of molecular spins up and down.} 
}
\end{figure}

Finally, we come to the main point of the work and discuss how population of collective rotational states can be measured in a molecular array interacting with a Rydberg superatom. 
The measurement is based on the interaction induced splittings between spin groups $\ket{k\uparrow,(N-k)\downarrow}$ for $k=0,...N$, analyzed in the previous section. 
As shown in Fig.\ref{fig:many-body-fluorescence}, first, the combined system is excited selectively 
from some group of $k \uparrow$ spins up $\ket{g,g,...g}\ket{\Psi_{\rm mol\;k \uparrow}}$ to the 
blockaded state $\ket{\Psi_{\rm atom}}\ket{\Psi_{\rm mol\;k \uparrow}}$ by a $\pi$ pulse, followed by another $\pi$ pulse connecting the 
$\ket{\Psi_{\rm atom}}$ state to some 
$\ket{\Psi_{e}}=\frac{1}{\sqrt{N_{a}}}\sum_{j=1}^{N_{a}}e^{i(k_{x\;{\rm Rydb}}-k_{x\;e})X_{j}}\ket{g_{1},g_{2},...,e_{j},...,g_{N_{a}}}$ state, where $\ket{e}$ is an atomic state rapidly decaying 
to $\ket{g}$. Again, instead of two $\pi$ pulses a sequence of either STIRAP or ARP pulses can be used. By repeating these excitation-fluorescence cycles the 
population of the $\ket{\Psi_{\rm mol\;k \uparrow}}$ can be detected. Let us illustrate the scheme for $N=3$ molecules. Initially, the molecular system is in a 
superposition of all spin states: 
\begin{eqnarray}
\ket{\Psi_{\rm mol}}=a_{\downarrow,\downarrow,\downarrow}\ket{\downarrow,\downarrow,\downarrow}+a_{\uparrow,\downarrow,\downarrow}\ket{\uparrow,\downarrow,\downarrow}+ \nonumber \\
+a_{\downarrow,\uparrow,\downarrow}\ket{\downarrow,\uparrow,\downarrow}+a_{\downarrow,\downarrow,\uparrow}\ket{\downarrow,\downarrow,\uparrow} + \nonumber \\
+a_{\uparrow,\uparrow,\downarrow}\ket{\uparrow,\uparrow,\downarrow}+a_{\uparrow,\downarrow,\uparrow}\ket{\uparrow,\downarrow,\uparrow}+ \nonumber \\
+a_{\downarrow,\uparrow,\uparrow}\ket{\downarrow,\uparrow,\uparrow}+a_{\uparrow,\uparrow,\uparrow}\ket{\uparrow,\uparrow,\uparrow}.  \nonumber
\end{eqnarray}
Suppose one is to measure the total population of the states with one spin up and two spins down, i.e. the $\ket{\uparrow,\downarrow,\downarrow}$, 
$\ket{\downarrow,\uparrow,\downarrow}$,  $\ket{\downarrow,\downarrow,\uparrow}$ states. For that the initial state $\ket{g,g,...,g}\ket{\Psi_{\rm mol}}$ is 
transformed into 
\begin{eqnarray*}
\lefteqn{\ket{g,g,...,g}\left(a_{\downarrow,\downarrow,\downarrow}\ket{\downarrow,\downarrow,\downarrow}+a_{\uparrow,\uparrow,\downarrow}\ket{\uparrow,\uparrow,\downarrow}+ \right. }\\
\lefteqn{\quad \left. +a_{\uparrow,\downarrow,\uparrow}\ket{\uparrow,\downarrow,\uparrow}+a_{\downarrow,\uparrow,\uparrow}\ket{\downarrow,\uparrow,\uparrow}+ \right. }\\
\lefteqn{\quad\left. +a_{\uparrow,\uparrow,\uparrow}\ket{\uparrow,\uparrow,\uparrow}\right)+ \ket{\Psi_{\rm atom}}\left(a_{\uparrow,\downarrow,\downarrow}\ket{\uparrow,\downarrow,\downarrow}+ \right. } \\
\lefteqn{\quad\left. +a_{\downarrow,\uparrow,\downarrow}\ket{\downarrow,\uparrow,\downarrow}+a_{\downarrow,\downarrow,\uparrow}\ket{\downarrow,\downarrow,\uparrow} \right)\;=}  \\
&=&\ket{g,g,...,g}\left(\ket{\Psi_{\rm mol}}-\ket{\Psi_{\rm mol\;k\uparrow=1}}\right)+\ket{\Psi_{\rm atom}}\ket{\Psi_{\rm mol\;k\uparrow=1}}, 
\end{eqnarray*}
by selective excitation to the Rydberg superatom state, 
where we denote the part of the state corresponding to a single spin up as 
\begin{eqnarray}
\ket{\Psi_{\rm mol\;k \uparrow=1}}=a_{\uparrow,\downarrow,\downarrow}\ket{\uparrow,\downarrow,\downarrow}+ a_{\downarrow,\uparrow,\downarrow}\ket{\downarrow,\uparrow,\downarrow}+a_{\downarrow,\downarrow,\uparrow}\ket{\downarrow,\downarrow,\uparrow}. \nonumber 
\end{eqnarray}
Next, atoms in $\ket{\Psi_{\rm atom}}$ are transferred to the state $\ket{\Psi_{e}}$, which rapidly decays to the $\ket{g,g,...,g}$ state. Repeating the 
excitation-fluorescence cycles and detecting fluorescence intensity will allow one to obtain the total population of the $\ket{\Psi_{\rm mol\;k \uparrow=1}}$ state, given by 
$|a_{\uparrow,\downarrow,\downarrow}|^{2}+|a_{\downarrow,\uparrow,\downarrow}|^{2}+|a_{\downarrow,\downarrow,\uparrow}|^{2}$.  
Applying this sequence for all $\ket{\Psi_{\rm mol\;k \uparrow}}$, 
 populations of all spin groups can be measured. This measurement is non-destructive with respect to molecules and is also of a QND 
type and will in fact project the molecular system to a superposition of states with a certain number of spins up and down provided a spontaneously 
emitted photon is detected. In this way entangled many-body molecular states can be prepared by measurement, which does not require molecules to interact, 
similar to 
the proposals in cQED systems \cite{Entanglement-by-measur-cQED}. For example, 
if initially all molecules are prepared in an equal superposition state $\left(\ket{\downarrow}_{i}+\ket{\uparrow}_{i}\right)/\sqrt{2}$, and the described above measurement sequence 
is applied, one will be able to project the system to an entangled state, which is an equal superposition of states with one spin up - W state, and more 
generally, equal superpositions of states with $k$ spins up and $N-k$ spins down - Dicke states. The collective states readout 
scheme could also be applied to measure the estimated Hamming weight of the molecular spin string $N_{\rm est}=\sum k \uparrow p_{k \uparrow}$, 
where $0 \le k \uparrow \le N$ is the number of spins 
up in a particular spin group and $p_{k \uparrow}$ is the probability of such a group, 
measured in our case by atomic fluorescence intensity. The Hamming weight, which is
 the total number of spins up in a string of $N$ spins or qubits, is a usefull quantuity in quantum error correction 
\cite{Error-corr} and in ion string clocks for determining the 
deviation of the clock frequency from an unperturbed ion frequency \cite{Ion-clock-Hamming-weight}.

\section{Conclusion}

We present a detailed analysis for non-destructive readout of mesoscopic ensembles of polar molecules, by exploiting the exquisite sensitivity of 
Rydberg states in interaction with molecular rotational states. The extreme dipole moment of the 
Rydberg atoms allows selective addressing of single or collective molecular rotational states. Our earlier proposal dealt only with single 
atoms and single molecules and found that, for example for distances of 300 - 600 nm between molecule and atom a  shift of several MHz 
can be detected in the Rydberg level depending on the molecular rotational state. This is wider than any line widths in this setup. 

In the 
present article, we have shown that these shifts, and the ensuing possibility of conditional Rydberg excitation and thus atom-molecule entanglement allows 
non-demolition readout not only for single atom-molecule pairs but also for collective rotational states in molecular ensembles. In the latter case 
instead of a single atom a 
Rydberg superatom, i.e., a single Rydberg excitation of a small ensemble of atoms, can be used.

In particular, detailed numerical estimates for small arrays of ground state KRb or RbYb molecules show that the difference of one excited 
collective rotational state leads to shifts of 100s of kHz in a Rydberg superatom about half a $\mu$m away. Our calculations were 
done for atoms in both a linear array and a dipole trap, often an easier experimental setup. 
Calculations in this case predict only a slightly smaller shift, which shows that the regularity in an optical lattice is not the defining 
feature of the setup.

While, at the present status of experiments, arrays of different species of under a $\mu$m distance might still be challenging (although two
close lattices were demostrated for the same atomic species \cite{Greiner-parallel-lattices}), 
the techniques presented here address particularly the non-destructive 
 readout of single or collective molecular rotational states, which 
has been a mostly unsolved problem to date. With the size of the Rydberg shifts of hundreds kHz, the superatom can be conditionally excited depending on 
the collective molecular state. While the fluorescence of the atom(s) is measured, effectively reading out the molecular state, the molecules 
remain untouched. An obvious extension would be to use the Rydberg atom(s) as a communication channel between two molecules or two molecular ensembles, thus 
allowing for effective indirect interactions between the molecular dipoles that are much stronger than the direct dipole-dipole 
interaction between the molecules. Proposals for many-body states based on strongly interacting dipoles could potentially be realized much 
easier this way. In addition, the conditional excitation allows very well controlled entanglement between molecules or groups of molecules, 
thus opening the door to, for example, entanglement-enhanced metrology.

\section{Acknowledgements}

E.K. was supported by the Russian Science Foundation (grant 16-12-00028) in the part of analysis of rotational states readout via 
atom-molecule entanglement, and the Russian Fund for Basic 
Research (grants RFBR 14-02-00174,16-52-10088-KO-a), and would like to thank the Institute for Theoretical Atomic and Molecular 
Physics for hospitality and financial 
support during her visit. SFY would like to thank the NSF for funding.

\section{Appendix A}

\begin{widetext}

In order 
to calculate the matrix elements of the interaction term $(V_{\rm e^--M})_{nlmJm_{J},n'l'm'J'm_{J}'}=\bra{J,m_{J}} e \vec{d}\ket{J',m_{J}'}\bra{n\;l\;m}\frac{\vec{R}}{R^{3}}\ket{n'\;l'\;m'}-\bra{J,m_{J}} e \vec{d} \ket{J',m_{J}'}\bra{n\;l\;m}\frac{\vec{R}-\vec{r}}{|\vec{R}-\vec{r}|^{3}}\ket{n'\;l'\;m'}$ 
we assume that the Rydberg atom is fixed in its position by e.g. trapping in a deep dipole trap or optical lattice such that $\vec{R}=R\vec{e}_{z}$. 
The molecular dipole moment can be written as $\vec{d}=d_{x}\vec{e}_{x}+d_{y}\vec{e}_{y}+d_{z}\vec{e}_{z}$ leading to:
\begin{eqnarray}
\vec{d}(\vec{R}-\vec{r})=d_{z}(R-r\cos \theta)-d_{x}r\sin \theta \cos \phi -d_{y}r\sin \theta \sin \phi,
\end{eqnarray} 
where $\theta$ and $\phi$ are the polar and azimuthal angles of the electron with respect to the Rydberg core.

As a result, 
\begin{eqnarray}
V_{\rm e^--M}=\frac{ed_{z}}{R^{2}}-\frac{ed_{z}(R-r\cos \theta)}{\left(R^{2}+r^{2}-2rR\cos \theta\right)^{3/2}}+\frac{ed_{x}r\sin \theta \cos \phi+d_{y}r\sin \theta \sin \phi}{\left(R^{2}+r^{2}-2rR\cos \theta\right)^{3/2}},
\end{eqnarray}
Next we will analyze separately the 
three terms of $V_{\rm e^--M}$:

1) The matrix element of the interaction between the molecular dipole and the Rydberg core has the form:

\begin{equation}
(V_{\rm e^--M}^{\rm core})_{nlmJm_{J},n'l'm'J'm_{J}'}=\bra{J,m_{J}}\bra{n\;l\;m}\frac{ed_{z}}{R^{2}}\ket{J'm_{J}'}\ket{n'\;l'\;m'}=\frac{ed_{z}^{J,m_{J};J',m_{J}'}}{R^{2}}\delta_{n,n'}\delta_{l,l'}\delta_{m,m'}\delta_{J\pm 1,J'}\delta_{m_{J},m_{J}'},
\end{equation}

2) Second contribution can be written in the following form:
\begin{eqnarray}
V_{\ e^--M}^{I} & = & -\frac{d_{z}(R-r\cos \theta)}{\left(R^{2}+r^{2}-2rR\cos \theta\right)^{3/2}}=d_{z}\frac{d}{dR}\frac{1}{\left(R^{2}+r^{2}-2rR\cos \theta\right)^{1/2}}= \\
&& =d_{z}\frac{d}{dR}\left \{\begin{array}{ccc} \sum_{l''=0}^{\infty}\sqrt{\frac{4\pi}{2l''+1}}Y_{l''0}(\theta, \phi)\frac{r^{l''}}{R^{l''+1}} & {\rm for}  & r<R  \nonumber \\
                                              \sum_{l''=0}^{\infty}\sqrt{\frac{4\pi}{2l''+1}}Y_{l''0}(\theta, \phi)\frac{R^{l''}}{r^{l''+1}} & {\rm for} & r>R 
\end{array} = \right. \nonumber \\
&& =d_{z}\left \{\begin{array}{ccc} -\sum_{l''=0}^{\infty}(l''+1)\sqrt{\frac{4\pi}{2l''+1}}Y_{l''0}(\theta,\phi)\frac{r^{l''}}{R^{l''+2}} & {\rm for} & r<R  \nonumber \\
                                  \sum_{l''=0}^{\infty}l''\sqrt{\frac{4\pi}{2l''+1}}Y_{l''0}(\theta,\phi)\frac{R^{l''-1}}{r^{l''+1}} & {\rm for} & r>R \nonumber 
\end{array} \right.
\end{eqnarray}

3) The third contribution can be written as follows: 
\begin{eqnarray}
V_{\rm e^--M}^{II}=\frac{d_{x}r\sin \theta \cos \phi+d_{y}r \sin \theta \sin \phi}{\left(R^{2}+r^{2}-2rR\cos \theta \right)^{3/2}}= 
\frac{r}{2}\sin \theta \frac{(d_{x}-id_{y})e^{i\phi}+(d_{x}+id_{y})e^{-i\phi}}{\left(R^{2}+r^{2}-2rR\cos \theta\right)^{3/2}}= \nonumber \\
=-\frac{1}{2R}\left((d_{x}-id_{y})e^{i\phi}+(d_{x}+id_{y})e^{-i\phi}\right)\frac{d}{d\theta}\frac{1}{\left(R^{2}+r^{2}-2rR\cos \theta\right)^{1/2}}
\end{eqnarray}
Let us use the fact that 
\begin{eqnarray}
\frac{1}{\left(R^{2}+r^{2}-2rR\cos \theta \right)^{1/2}}=\left \{\begin{array}{ccc} \sum_{l''=0}^{\infty}P_{l''}(\cos \theta)\frac{r^{l''}}{R^{l''+1}} & {\rm for} & r<R \nonumber \\
                                              \sum_{l''=0}^{\infty}P_{l''}(\cos \theta)\frac{R^{l''}}{r^{l''+1}} & {\rm for} & r>R 
\end{array} \right.
\end{eqnarray}
and
\begin{eqnarray}
\frac{d}{d\theta}P_{l''}(\cos \theta)e^{\pm i\phi}=\left \{\begin{array}{cc}
                                                             P_{l''}^{1}(\cos \theta)e^{i\phi}=\sqrt{\frac{4\pi l''(l''+1)}{2l''+1}}Y_{l''}^{1}(\theta,\phi)  \nonumber \\
                                                             -l''(l''+1)P_{l''}^{-1}(\cos \theta)e^{-i\phi}=-\sqrt{\frac{4\pi l''(l''+1)}{2l''+1}}Y_{l''}^{-1}(\theta,\phi)
\end{array} \right.
\end{eqnarray}

As a result, the third term becomes:
\begin{eqnarray}
V_{\rm e^--M}^{II}=\frac{d_{x}r\sin \theta \cos \phi+d_{y}r \sin \theta \sin \phi}{\left(R^{2}+r^{2}-2rR\cos \theta \right)^{3/2}}= \\
=-\frac{d_{x}-id_{y}}{2R}\left \{\begin{array}{ccc} \sum_{l''=0}^{\infty}\frac{r^{l''}}{R^{l''+1}}\sqrt{\frac{4\pi l''(l''+1)}{2l''+1}}Y_{l''}^{1}(\theta,\phi) & {\rm for} & r<R \nonumber \\
                                                    \sum_{l''=0}^{\infty}\frac{R^{l''}}{r^{l''+1}}\sqrt{\frac{4\pi l''(l''+1)}{2l''+1}}Y_{l''}^{1}(\theta,\phi) & {\rm for} & r>R 
\end{array}\right. \nonumber \\
+ \frac{d_{x}+id_{y}}{2R}\left \{\begin{array}{ccc} \sum_{l''=0}^{\infty}\frac{r^{l''}}{R^{l''+1}}\sqrt{\frac{4\pi l''(l''+1)}{2l''+1}}Y_{l''}^{-1}(\theta,\phi) & {\rm for} & r<R \nonumber \\
                                                    \sum_{l''=0}^{\infty}\frac{R^{l''}}{r^{l''+1}}\sqrt{\frac{4\pi l''(l''+1)}{2l''+1}}Y_{l''}^{-1}(\theta,\phi) & {\rm for} & r>R 
\end{array}\right.
\end{eqnarray}

The above expansions allow one to calculate the matrix elements of the $V_{\rm e^--M}^{I}$ and $V_{\rm e^--M}^{II}$ terms:
\begin{eqnarray}
\lefteqn{\bra{J,m_{J}}\bra{n\;l\;m} V_{\rm e^--M}^{I} \ket{n'\;l'\;m'}\ket{J',m_{J}'}\;= }\\
&=&\delta_{m,m'}\left(-\sum_{l''=0}^{\infty}(l''+1)\sqrt{\frac{4\pi}{2l''+1}}\frac{1}{R^{l''+2}}\int_{0}^{R}r^{l''+2}R_{nl}(r)R_{n'l'}(r)dr\int_{0}^{2\pi}d\phi \int_{0}^{\pi} Y_{l}^{m*}Y_{l''}^{0}Y_{l'}^{m'}\sin \theta d\theta + \right. \nonumber \\
&& \left. +\sum_{l''=0}^{\infty}l''\sqrt{\frac{4\pi}{2l''+1}}R^{l''-1}\int_{R}^{\infty}\frac{1}{r^{l''-1}}R_{nl}(r)R_{n'l'}(r)dr\int_{0}^{2\pi}d\phi\int_{0}^{\pi} Y_{l}^{m*}Y_{l''}^{0}Y_{l'}^{m'}\sin \theta d\theta \right)\bra{J,m_{J}}ed_{z}\ket{J',m_{J}'}= \nonumber \\
&=&\delta_{m,m'}\delta_{J',J\pm 1}\delta_{m_{J},m_{J}'}\left(-\sum_{l''=0}^{\infty}(l''+1)\sqrt{\frac{4\pi}{2l''+1}}\frac{1}{R^{l''+2}}\int_{0}^{R}r^{l''+2}R_{nl}(r)R_{n'l'}(r)dr\int_{0}^{\pi} Y_{l}^{m*}Y_{l''}^{0}Y_{l'}^{m}\sin \theta d\theta  + \right. \nonumber \\
&&\left. +\sum_{l''=0}^{\infty}l''\sqrt{\frac{4\pi}{2l''+1}}R^{l''-1}\int_{R}^{\infty}\frac{1}{r^{l''-1}}R_{nl}(r)R_{n'l'}(r)dr\int_{0}^{\pi} Y_{l}^{m*}Y_{l''}^{0}Y_{l'}^{m}\sin \theta d\theta \right)ed_{z}^{J,m_{J};J',m_{J}'}\nonumber
\end{eqnarray}
Next,
\begin{eqnarray*}
\lefteqn{\bra{J,m_{J}}\bra{n\;l\;m}V_{\rm e^--M}^{II}\ket{n'\;l'\;m'}\ket{J',m_{J}'}\;=}  \\
&=&-\left(\sum_{l''=0}^{\infty}\frac{1}{R^{l''+1}}\sqrt{\frac{4\pi l''(l''+1)}{2l''+1}}\int_{0}^{R}r^{l''+2}R_{nl}(r)R_{n'l'}dr\int_{0}^{2\pi} d\phi \int_{0}^{\pi} Y_{l}^{m*}Y_{l''}^{1}Y_{l'}^{m'}\sin \theta d\theta + \right. \\
&&\left. + \sum_{l''=0}^{\infty}R^{l''}\sqrt{\frac{4\pi l''(l''+1)}{2l''+1}}\int_{R}^{\infty}\frac{R_{nl}R_{n'l'}}{r^{l''-1}}dr\int_{0}^{2\pi} d\phi \int_{0}^{\pi} Y_{l}^{m*}Y_{l''}^{1}Y_{l'}^{m'}\sin \theta d\theta \right)\frac{\bra{J,m_{J}}d_{x}-id_{y}\ket{J',m_{J}'}}{2R} \nonumber \\
&&+\left(\sum_{l''=0}^{\infty}\frac{1}{R^{l''+1}}\sqrt{\frac{4\pi l''(l''+1)}{2l''+1}}\int_{0}^{R}r^{l''+2}R_{nl}(r)R_{n'l'}dr\int_{0}^{2\pi } d\phi \int_{0}^{\pi} Y_{l}^{m*}Y_{l''}^{-1}Y_{l'}^{m'}\sin \theta d\theta  + \right. \\
&&\left. + \sum_{l''=0}^{\infty}R^{l''}\sqrt{\frac{4\pi l''(l''+1)}{2l''+1}}\int_{R}^{\infty}\frac{R_{nl}R_{n'l'}}{r^{l''-1}}dr\int_{0}^{2\pi} d\phi \int_{0}^{\pi} Y_{l}^{m*}Y_{l''}^{-1}Y_{l'}^{m'}\sin \theta d\theta \right)\frac{\bra{J,m_{J}} d_{x}+id_{y} \ket{J',m_{J}'}}{2R}\;= \\
&=&-\frac{d_{+}^{J,m_{J};J',m_{J}'}}{\sqrt{2}R}\delta_{m,m'+1}\left(\sum_{l''=0}^{\infty}\frac{1}{R^{l''+1}}\sqrt{\frac{4\pi l''(l''+1)}{2l''+1}}\int_{0}^{R}r^{l''+2}R_{nl}(r)R_{n'l'}dr\int Y_{l}^{m*}Y_{l''}^{1}Y_{l'}^{m'}\sin \theta' d\theta' d\phi' + \right.  \\
&&\left. + \sum_{l''=0}^{\infty}R^{l''}\sqrt{\frac{4\pi l''(l''+1)}{2l''+1}}\int_{R}^{\infty}\frac{R_{nl}R_{n'l'}}{r^{l''-1}}dr\int Y_{l}^{m*}Y_{l''}^{1}Y_{l'}^{m'}\sin \theta' d\theta' d\phi'\right)+ \\
&&-\frac{d_{-}^{J,m_{J};J',m_{J}'}}{\sqrt{2}R}\delta_{m,m'-1}\left(\sum_{l''=0}^{\infty}\frac{1}{R^{l''+1}}\sqrt{\frac{4\pi l''(l''+1)}{2l''+1}}\int_{0}^{R}r^{l''+2}R_{nl}(r)R_{n'l'}dr\int Y_{l}^{m*}Y_{l''}^{-1}Y_{l'}^{m'}\sin \theta' d\theta' d\phi' + \right. \\
&&\left. + \sum_{l''=0}^{\infty}R^{l''}\sqrt{\frac{4\pi l''(l''+1)}{2l''+1}}\int_{R}^{\infty}\frac{R_{nl}R_{n'l'}}{r^{l''-1}}dr\int Y_{l}^{m*}Y_{l''}^{-1}Y_{l'}^{m'}\sin \theta' d\theta' d\phi'\right), 
\end{eqnarray*}
where the dipole moment matrix elements are $d_{z}^{Jm_{J},J'm_{J}'}=\bra{J,m_{J}}d_{z}\ket{J',m_{J}'}$, 
$d_{\pm}^{Jm_{J},J'm_{J}'}=\pm \bra{J,m_{J}}d_{x}\mp id_{y}\ket{J',m_{J}'}/\sqrt{2}$, where for $J=0$, $J=1$ and $J=2$ rotational states the corresponding matrix 
elements are $d_{z}^{0,0;1,0}=d/\sqrt{3}$, $d_{z}^{1,0;2,0}=2d/\sqrt{15}$, $d_{z}^{1,\pm 1;2,\pm 1}=d/\sqrt{5}$, $d_{\pm}^{0,0;1,\pm 1}=-d/\sqrt{3}$, $d_{\pm}^{1,0;2,\pm 1}=-d/\sqrt{5}$, 
$d_{\pm}^{1,\pm 1;2,\pm 2}=-d\sqrt{2}/\sqrt{5}$, $d_{\pm}^{1,\pm 1;2,0}=-d/\sqrt{15}$, where $d$ is the permanent dipole moment of a molecule.  
The integrals involving three spherical harmonics are calculated using the expression:

\begin{eqnarray*}
\lefteqn{\int_{0}^{2\pi}\int_{0}^{\pi}Y_{l_{1}}^{m_{1}}(\theta,\phi)Y_{l_{2}}^{m_{2}}(\theta,\phi)Y_{l_{3}}^{m_{3}}(\theta,\phi)\sin \theta d\theta d\phi\;= }\\
&=&=\sqrt{\frac{(2l_{1}+1)(2l_{2}+1)(2l_{3}+1)}{4\pi}}\left(\begin{array}{ccc} l_{1} & l_{2} & l_{3} \\
                                                                            0 & 0 & 0 
                                                                            \end{array} \right)
                                                   \left(\begin{array}{ccc} l_{1} & l_{2} & l_{3} \\
                                                                            m_{1} & m_{2} & m_{3} 
                                                                            \end{array} \right). 
\end{eqnarray*}

The matrix elements of the full Hamiltonian including all terms are then given by 
\begin{eqnarray}
\lefteqn{\bra{J,m_{J}}\bra{n\;l\;m}H\ket{n'\;l'\;m'}\ket{J',m_{J}'}\;=} \\
&=&-\frac{1}{2(n-\mu_{l})^{2}}\delta_{n,n'}\delta_{l,l'}\delta_{m,m'}\delta_{J,J'}\delta_{m_{J},m_{J}'}
+BJ(J+1)\delta_{n,n'}\delta_{l,l'}\delta_{m,m'}\delta_{J,J'}\delta_{m_{J},m_{J}'} \nonumber \\
&&+\bra{J,m_{J}}\bra{n\;l\;m}V_{\rm e^--M}^{\rm core}\ket{n'\;l'\;m'}\ket{J',m_{J}'}+\bra{J,m_{J}}\bra{n\;l\;m}V_{\rm e^--M}^{I}\ket{n'\;l'\;m'}\ket{J',m_{J}'}\nonumber \\
&&+\bra{J,m_{J}}\bra{n\;l\;m}V_{\rm e^--M}^{II}\ket{n'\;l'\;m'}\ket{J',m_{J}'}. \nonumber
\end{eqnarray}
The total Hamiltonian is then diagonalized to find new eigenstates accounting for the interaction and corresponding eigenenergies.

\section{Appendix B}

Let us consider the setup shown in Fig.\ref{fig:setup}b where a molecule is situated at a distance $\Delta x$ with respect to the normal connecting 
the X axis and the Rydberg core. in this case $\vec{R}=\frac{\Delta x}{R}\vec{e}_{X}+\frac{\rho}{R}\vec{e}_{Z}$, and the analysis of Appendix A, which was 
based on the assumption that $\vec{R}=R\vec{e}_{Z}$ is not directly applicable. However, if one transforms the coordinate system from (X,Y,Z) to (x,y,z) such 
that in the new system $\vec{R}=R\vec{e}_{z}$, i.e. parallel to the new z axis, the analysis of Appendix A is valid again with one correction. The coordinate 
transformation is equivalent to the rotation of the system with respect to Y axis by an angle $\beta=-\arcsin\left(\Delta x/R\right)$. Such a rotation leads to 
transformation of molecular rotational states as $\ket{J,m_{J}}=\sum_{m_{J}'}D^{(J)}_{m_{J}'m_{J}}(\alpha,\beta,\gamma)\ket{J,m_{J}'}$, where $\alpha$, $\beta$ and 
$\gamma$ are Euler's angles (in our case $\alpha=\gamma=0$), $\ket{J,m_{J}}$ and $\ket{J,m_{J}'}$ are rotational states before and after the rotation, and $J$ is conserved during the 
transformation. For $J=0$ the state does not change $\ket{J=0,m_{J}=0}=\ket{J=0,m_{J}'=0}$, and for $J=1$ the states transform as follows \cite{Landavshits}:
\begin{eqnarray}
\ket{J=1,m=0}&=&\frac{1}{\sqrt{2}}\sin \beta \ket{J'=1,m_{J}'=1}+\cos \beta \ket{J'=1,m_{J}'=0}-\frac{1}{\sqrt{2}}\sin \beta \ket{J'=1,m_{J}'=-1}, \\
\ket{J=1,m=-1}&=&\frac{1}{2}\left(1-\cos \beta \right) \ket{J'=1,m_{J}'=1}+\frac{1}{\sqrt{2}}\sin \beta \ket{J'=1,m_{J}'=0}+\frac{1}{2}\left(1+\cos \beta \right)\ket{J'=1,m_{J}'=-1}, \nonumber \\
\ket{J=1,m=1}&=&\frac{1}{2}\left(1+\cos \beta \right) \ket{J'=1,m_{J}'=1}-\frac{1}{\sqrt{2}}\sin \beta \ket{J'=1,m_{J}'=0}+\frac{1}{2}\left(1-\cos \beta \right)\ket{J'=1,m_{J}'=-1}, \nonumber
\end{eqnarray}
Matrix elements of $V_{\rm e^--M}$ between $\ket{J',m_{J}'}$ in the transformed system can now be calculated as in Appendix A.

\end{widetext}


\begin{thebibliography}{11}

\bibitem{QC-pol-mol} D. DeMille, Phys. Rev. Lett. {\bf 88}, 067901 (2002); 
A. Andre, D. DeMille, J. M. Doyle, M. D. Lukin, S. E. Maxwell, P. Rabl, R. J. Shoelkopf, P. Zoller, Nat. Phys. 
{\bf 2}, 636 (2006); 
S. Yelin, K. Kirby, R. Cote, 
Phys. Rev. Lett. {\bf 74}, 050301 (2006).   

\bibitem{Pol-mol-quant-magn} A. V. Gorshkov, S. R. Manmana, G. Chen, J. Ye, E. Demler, M. D. Lukin, A. M. Rey, Phys. Rev. Lett. {\bf 107}, 115301 (2011).

\bibitem{Pol-mol-topology} S. R. Manmana, E. M. Stoudenmire,  K. R. A. Hazzard,  A. M. Rey,  A. V. Gorshkov, Phys. Rev. A {\bf 87}, 081106(R) (2013).

\bibitem{Pol-mol-entanglement} K. R. A. Hazzard, M. van den Worm, M. Foss-Feig, S. R. Manmana, E. G. Dalla Torre, T. Pfau, M. Kastner, A. M. Rey, Phys. Rev. A 
{\bf 90}, 063622 (2014).

\bibitem{Jin-pol-mol-spin-exchange}  B. Yan, S. A. Moses, B. Gadway, J. P. Covey, K. R. A. Hazzard, A. M. Rey, D. S. Jin, J. Ye, Nature {\bf 501}, 521 (2013). 

\bibitem{Magnetic-spin-exchange} A. de Paz, A. Sharma, A. Chotia, E. Marechal, J. H. Huckans, P. Pedri, L. Santos, O. Gorceix, L. Vernac, B. Laburthe-Tolra, 
Phys. Rev. Lett. {\bf 111}, 185305 (2013).

\bibitem{REMPI} W. C. Stwalley, H. Wang, J. Mol. Spectr. {\bf 195}, 194 (1999).

\bibitem{PCCP} E. Kuznetsova, S. T. Rittenhouse, H. R. Sadeghpour, S. F. Yelin, Phys. Chem. Chem. Phys. {\bf 13}, 17115 (2011).

\bibitem{Seth-Hossein-PRL} S. T. Rittenhouse, H. R. Sadeghpour, Phys. Rev. Lett. {\bf 104}, 243002 (2010); 

\bibitem{Seth-JPB} S. T. Rittenhouse, M. Mayle, P. Schmelcher, H. R. Sadeghpour, J. Phys. B {\bf 44}, 184005 (2011).


\bibitem{Quantum-logic-spectroscopy} P. O. Schmidt, T. Rosenband, C. Langer, W. M. Itano, J. C. Bergquist, D. J. Wineland, 
Science {\bf 309}, 749 (2005); T. R. Tan, J. P. Gaebler, Y. Lin, Y. Wan, R. Bowler, D. Leibfried, D. J. Wineland, Nature {\bf 528}, 380 (2015); 
J. Mur-Petit, J. J. Garcia-Ripoll, J. Perez-Rios, J. Campos-Martinez, M. I. Hernandez, S. Willitsch, Phys. Rev. A {\bf 85}, 022308 (2012).

\bibitem{Garcia-Ripoll-pol-mol-ion} J. Mur-Petit, J. J. Garcia-Ripoll, Phys. Rev. A {\bf 91}, 012504 (2015); J. Mur-Petit, J. J. Garcia-Ripoll, Appl. Phys. B {\bf 114}, 283 (2014). 

\bibitem{QND} V. B. Braginsky, F. Ya. Khalili, Rev. Mod. Phys. {\bf 68}, 1, (1996).


\bibitem{N-V-nuclear-spin-control} L. Childress, M. V. Gurudev Dutt, J. M. Taylor, A. S. Zibrov, F. Jelezko, J. Wrachtrup, P. R. Hemmer, M. D.
Lukin, Science {\bf 314}, 281 (2006).


\bibitem{QDot-nuclear-spin-control} J. R. Petta, J. M. Taylor, A. C. Johnson, A. Yacoby, M. D. Lukin, C. M. Marcus, M. P. Hanson,A. C. Gossard,
Phys. Rev. Lett. {\bf 100}, 067601 (2008).

\bibitem{Nuclear-spin-polarization} B. Urbaszek, X. Marie, T. Amand, O. Krebs, P. Voisin, P. Maletinsky, A. Hogele, A. Imamoglu, Rev. Mod. Phys. 
{\bf 85}, 79 (2013).

\bibitem{Nuclear-spin-superradiance} E. M. Kessler, S. Yelin, M. D. Lukin, J. I. Cirac, G. Giedke, Phys. Rev. Lett. {\bf 104}, 143601 (2010).

\bibitem{N-V-nuclear-spin-squeezing} M. S. Rudner, L. M. K. Vandersypen, V. Vuletic, L. S.
Levitov, Phys. Rev. Lett. {\bf 107}, 206806 (2011).

\bibitem{Paola-N-V-envir-assist-measurement} G. Goldstein, P. Cappellaro, J. R. Maze, J. S. Hodges,
L. Jiang, A. S. Sorensen, M. D. Lukin, Phys. Rev. Lett. {\bf 106}, 140502 (2011).

\bibitem{Pfau-Rydb-BEC-spectroscopy} M. Schlagmller, T. C. Liebisch, H. Nguyen, G. Lochead, F. Engel, F. Bttcher, K. M. Westphal, K. S. Kleinbach,
R. Lw, S. Hofferberth, T. Pfau, Phys. Rev. Lett. {\bf 116}, 053001 (2016); R. Schmidt, H. R. Sadeghpour, E. Demler, Phys. Rev. Lett. {\bf 116}, 
105302 (2016).

\bibitem{Rydb-mol-spectra-Pfau} A. Gaj, A.T. Krupp, J.B. Balewski, R. Low, S. Hofferberth, T. Pfau, Nat. Comm. {\bf 5}, 4546 (2014).

\bibitem{Rydb-quant-simulator}  H. Weimer, M. Mller, I. Lesanovsky, P. Zoller, H. P. Buchler, Nat. Phys. {\bf 6}, 382 (2010).


\bibitem{Greiner-parallel-lattices} P. M. Preiss, R. Ma, M. E. Tai, J. Simon, M. Greiner, Phys. Rev. A {\bf 91}, 041602 (2015).

\bibitem{Rydb-fine-structure} D. A. Anderson, S. A. Miller, G. Raithel, Phys. Rev. Lett. {\bf 112}, 163201 (2014).

\bibitem{KRb-dip-mom} K.-K. Ni, S. Ospelkaus, M. H. G. de Miranda, A. Pe'er A, B. Neyenhuis, J. J. Zirbel, S. Kotochigova, P. S. Julienne, D. S. Jin, J. Ye, Science
{\bf 322}, 231 (2008).


\bibitem{KRb-rot-const} K.-K. Ni, S. Ospelkaus, D. J. Nesbitt, J. Ye, D. S. Jin, Phys. Chem. Chem. Phys. {\bf 11}, 9626 (2009).

\bibitem{KRb-ground-state} K.-K. Ni, S. Ospelkaus, M. H. G. de Miranda, A. Peer, B. Neyenhuis, J. J. Zirbel, S. Kotochigova, P. S. Julienne, D. S. Jin, J. Ye, 
Science {\bf 322}, 231 (2008).

\bibitem{RbYb-dip-mom-rot-const} S. N. Tohme, M. Korek, Chem. Phys. {\bf 410}, 37 (2013).

\bibitem{RbYb-experim} F. Munchow, C. Bruni, M. Madalinski, A. Gorlitz, Phys. Chem. Chem. Phys. {\bf 13}, 18734 (2011). 

\bibitem{RbYb-hyperfine} M. Borkowski, P. S. Zuchowski, R. Ciurylo, P. S. Julienne, D. Kedziera, L. Mentel,
P. Tecmer, F. Munchow, C. Bruni, A. Gorlitz, Phys. Rev. A {\bf 88}, 052708 (2013).


\bibitem{RbYb-theory} L. K. Sorensen, S. Knecht, T. Fleig, C. M. Marian, J. Chem. Phys. {\bf 113}, 12607 (2009); D. A. Brue, J. M. Hutson, Phys. Rev. A {\bf 87}, 
052709 (2013); M. Tomza, R. Gonzalez-Ferez, C. P. Koch, R. Morzynski, Phys. Rev. Lett. {\bf 112}, 113201 (2014). 

\bibitem{Param-mol-QC} F. Herrera, Y. Cao, S. Kais, K. B. Whaley, New J. Phys. {\bf 16}, 075001 (2014); M. Karra, K. Sharma, B. Friedrich, S. Kais, D. Herschbach, arxiv:1601.02409.

\bibitem{Param-mol-QSim} A. Micheli, G. K. Brennen, P. Zoller, Nat. Phys. {\bf 2}, 341 (2006).


\bibitem{RbCs-dip-mom} S. Kotochigova, E. Tiesinga, J. Chem. Phys. {\bf 123}, 174304 (2005).

\bibitem{RbCs-rot-const} P. K. Molony, P. D. Gregory, Z. Ji, B. Lu, M. P. Köppinger, C. R. Le Sueur, C. L. Blackley, J. M. Hutson, S. L. Cornish, Phys. Rev. Lett. 
{\bf 113}, 255301 (2014). 

\bibitem{LiNa} J. Deiglmayr, M, Aymar, R. Wester, M. Weidemüller, O. Dulieu, J. Chem. Phys. {\bf 129}, 064309 (2008).

\bibitem{Alkali-metal-alkaline-earth} R. Guerout, M. Aymar, O. Dulieu, Phys. Rev. A {\bf 82}, 042508 (2010).

\bibitem{NaCa} G. Gopakumar, M. Abe, M. Hada1, M. Kajita, J. Chem. Phys. {\bf 140}, 224303 (2014). 

\bibitem{Seth-Hossein-Others-Science-2011} W. Li, T. Pohl, J. M. Rost, S. T. Rittenhouse, H. R. Sadeghpour, J. Nipper, B. Butscher, J. B. Balewski, V.
Bendkowsky, R. Lw, T. Pfau, Science {bf 334}, 1110 (2011).

\bibitem{Curl} R. F. Curl, Mol. Phys. {\bf 9}, 585 (1965).

\bibitem{Munchow-thesis} F. Munchow, PhD thesis, University of Dusseldorf (2012).


\bibitem{Rydb-decay-rates} Rb($60s$) decay rate has contributions from spontaneous emission with the rate $\Gamma^{\rm sp}_{60s} = 723$ Hz and from the interaction with black-body 
radiation at 300 K, which includes BBR induced decay with the rate $\Gamma^{\rm BBR,d}_{60s} = 0.675\Gamma^{\rm sp}_{60s}=488$ Hz, excitation to higher Rydberg
states with the rate $\Gamma^{\rm BBR,e}=0.583\Gamma^{\rm sp}_{60s}=421.5$ Hz, and ionization with the rate $\Gamma^{\rm BBR,ion}_{60s}=0.0155\Gamma^{\rm sp}_{60s}=11.2$ Hz, 
resulting in the total decay rate $\Gamma_{60s} =\Gamma^{\rm sp}_{60s}+\Gamma^{\rm BBR,d}_{60s}+\Gamma^{\rm BBR,e}_{60s}+\Gamma^{\rm BBR,ion}_{60s}\approx 1.644$ kHz 
(V. D. Ovsiannikov, I. L. Glukhov, E. A. Nikipelov, J. Phys. B {\bf 44}, 195010 (2011)).
Molecular rotational states can also decay due to interaction with black-body radiation, but with much smaller rates $\sim 10^{-6}$ Hz for the ground vibrational 
state of KRb at 300 K (S. Kotochigova, E. Tiesinga, P.S. Julienne, Eur. Phys. J. D {\bf 31}, 189 (2004)), spontaneous emission decay rates being even smaller.


\bibitem{Dipole-blockade} D. Jaksch, J. I. Cirac, P. Zoller, S. L. Rolston, R. Cote, M. D. Lukin, Phys. Rev. Lett. {\bf 85}, 2208 (2000).


\bibitem{Ring-lattice} A. De Pasquale, P. Facchi, Phys. Rev. A {\bf 80}, 032102 (2009); 
B. Olmos, R. Gonzalez-Ferez, I. Lesanovsky, Phys. Rev. Lett. {\bf 103}, 185302 (2009); B. Olmos, I. Lesanovsky, Phys. Rev. A {\bf 82}, 063404 (2010).

\bibitem{Robin-Rydb} K. Singer, J. Stanojevic, M. Weidemuller, R. Cote, J. Phys. B {\bf 38}, S295 (2005).

\bibitem{Many-body-description} In (a) and (b) thick solid black curve corresponds to the $\ket{\downarrow,\downarrow,\downarrow}$ state, 
dashed thick and thin red curves 
correspond to the $\ket{\uparrow,\downarrow,\downarrow}$, $\ket{\downarrow,\downarrow,\uparrow}$ and $\ket{\downarrow,\uparrow,\downarrow}$ states, respectively; 
dotted thick and thin green curves correspond to the $\ket{\uparrow,\downarrow,\uparrow}$ and $\ket{\uparrow,\uparrow,\downarrow}$, 
$\ket{\downarrow,\uparrow,\uparrow}$ states, respectively, and thick dash-dotted blue curve corresponds to the $\ket{\uparrow,\uparrow,\uparrow}$ state; 
In (c) and (d) thick solid black curve corresponds to the 
$\ket{\downarrow,\downarrow,\downarrow,\downarrow,\downarrow}$ state, dashed thick, medium and thin red curves 
correspond to the $\ket{\uparrow,\downarrow,\downarrow,\downarrow,\downarrow}$, $\ket{\downarrow,\downarrow,\downarrow,\downarrow,\uparrow}$ and $\ket{\downarrow,\uparrow,\downarrow,\downarrow,\downarrow}$, 
$\ket{\downarrow,\downarrow,\downarrow,\uparrow,\downarrow}$ and $\ket{\downarrow,\downarrow,\uparrow,\downarrow,\downarrow}$ states, respectively; 
dotted green curves in the order of reducing thickness correspond to the $\ket{\uparrow,\downarrow,\downarrow,\downarrow,\uparrow}$ and 
$\ket{\uparrow,\downarrow,\downarrow,\uparrow,\downarrow}$, $\ket{\downarrow,\uparrow,\downarrow,\downarrow,\uparrow}$
$\ket{\downarrow,\downarrow,\uparrow,\downarrow,\uparrow}$, $\ket{\uparrow,\downarrow,\uparrow,\downarrow,\downarrow}$ and 
$\ket{\downarrow,\downarrow,\downarrow,\uparrow,\uparrow}$, $\ket{\uparrow,\uparrow,\downarrow,\downarrow,\downarrow}$  and $\ket{\downarrow,\uparrow,\downarrow,\uparrow,\downarrow}$ 
and $\ket{\downarrow,\uparrow,\uparrow,\downarrow,\downarrow}$, $\ket{\downarrow,\downarrow,\uparrow,\uparrow,\downarrow}$ states, respectively; 
blue dash-dotted curves in the order of reducing thickness correspond to the 
$\ket{\uparrow,\uparrow,\downarrow,\downarrow,\uparrow}$, $\ket{\uparrow,\downarrow,\downarrow,\uparrow,\uparrow}$ and 
 $\ket{\uparrow,\downarrow,\uparrow,\downarrow,\uparrow}$ 
and $\ket{\uparrow,\uparrow,\downarrow,\uparrow,\downarrow}$, $\ket{\downarrow,\uparrow,\downarrow,\uparrow,\uparrow}$, 
$\ket{\uparrow,\downarrow,\uparrow,\uparrow,\downarrow}$, $\ket{\downarrow,\uparrow,\uparrow,\downarrow,\uparrow}$ 
and  $\ket{\uparrow,\uparrow,\uparrow,\downarrow,\downarrow}$, $\ket{\downarrow,\downarrow,\uparrow,\uparrow,\uparrow}$ and 
$\ket{\downarrow,\uparrow,\uparrow,\uparrow,\downarrow}$ states, respectively; thick, medium and 
thin orange curves correspond to $\ket{\uparrow,\uparrow,\downarrow,\uparrow,\uparrow}$ and $\ket{\uparrow,\uparrow,\uparrow,\downarrow,\uparrow}$, 
$\ket{\uparrow,\downarrow,\uparrow,\uparrow,\uparrow}$ and $\ket{\downarrow,\uparrow,\uparrow,\uparrow,\uparrow}$, $\ket{\uparrow,\uparrow,\uparrow,\uparrow,\downarrow}$ states, respectively, and thick pink dash-dot-doted curve corresponds to the 
$\ket{\uparrow,\uparrow,\uparrow,\uparrow,\uparrow}$ state.


\bibitem{Atom-microscope} C. Weitenberg, M. Endres, J. F. Sherson, M. Cheneau, P. Schauß, T. Fukuhara, I. Bloch, S. Kuhr, Nature {\bf 471}, 319 (2011).


\bibitem{Accordeon-lattice} R. A. Williams, J. D. Pillet, S. Al-Assam, B. Fletcher, M. Shotter, C. J. Foot, Opt. Expr. {\bf 16}, 16977 (2008); 
S. Al-Assam, R. A. Williams, C. J. Foot, Phys. Rev. A {\bf 82}, 021604 (2010).

\bibitem{STIRAP-excitation} Instead of a $\pi$ pulse one can excite the atom to the Rydberg state using STIRAP (K. Bergmann, H. Theuer, B. W. Shore, Rev. Mod. Phys. {\bf 70}, 1003 (1998)) or 
adiabatic rapid passage (ARP) techniques (E. Kuznetsova, G. Liu, S. Malinovskaya, Phys. Scr. T {\bf 160}, 014024 (2014)).
Their advantage over a $\pi$ pulse is that precise tuning of the pulse area to $\pi$ is not required, i.e. STIRAP/ARP 
excitation is robust with respect to pulse duration and Rabi frequency. An additional advantage 
of ARP excitation is that one does not have to know precisely the shifts of the $\ket{r}\ket{\uparrow}$ and $\ket{r}\ket{\downarrow}$ states, 
provided that the shifts have opposite signs. 


\bibitem{QND-theory} T. C. Ralph, S. D. Bartlett, J. L. O'Brien, G. J. Pryde, H. M. Wiseman, Phys. Rev. A {\bf 73}, 012113 (2006).



\bibitem{QND-ions} D. B. Hume, T. Rosenband, D. J. Wineland, Phys. Rev. Lett. {\bf 99}, 120502 (2007). 

\bibitem{QND-N-V} P. Neumann, J. Beck, M. Steiner, F. Rempp, H. Fedder, P. R. Hemmer, J. Wrachtrup, F. Jelezko, Science {\bf 329}, 542 (2010).

\bibitem{Beterov-Saffman-QLS} I. I. Beterov, M. Saffman, Phys. Rev. A {\bf 92}, 042710 (2015).


\bibitem{Entanglement-by-measur-cQED} F. Helmer, F. Marquardt, Phys. Rev. A {\bf 79}, 052328 (2009).

\bibitem{Error-corr} A. M. Steane, Proceedings of the International School of Physics "Enrico Fermi", edited by G. Casati, D. L. Shepelyansky, P. Zoller, 
p. 1-32 (IOS Press, Amsterdam, 2006).

\bibitem{Ion-clock-Hamming-weight} M. Schulte, N. Lorch, I. D. Leroux, P. O. Schmidt, K. Hammerer, arxiv:1501.06453.


\bibitem{Landavshits} L. D. Landau, E. M. Livshitz, {\it Quantum Mechanics}, Pergamon Press, 1965.

\end{thebibliography}
\end{document}